\documentclass[]{article}

\usepackage{subfigure,url}

\usepackage{graphicx}

\usepackage{amsmath}
\usepackage{amssymb}
\usepackage{verbatim}

\newcommand{\Phys}{\textit{Physarum }}

\def\degree{\kern-.2em\r{}\kern-.3em\xspace}
\def\degC{\kern-.2em\r{}\kern-.3em C\xspace}

\begin{document}

\title{Towards {\itshape Physarum} Engines}

\author{Soichiro Tsuda, Jeff Jones, and Andrew Adamatzky\\ \\Unconventional Computing Centre,\\ University of the
West of England, \\Bristol BS16 1QY, United Kingdom}
\thanks{Unconventional Computing Centre, University of the
West of England, Bristol BS16 1QY, United Kingdom}

\date{Published in: \\{\bf Tsuda~S., Jones~J., Adamatzky~A. \\Towards Physarum engines. \\ Applied Bionics and Biomechanics 9 (2012) 3, 221-240.}}

\maketitle

\begin{abstract}
The slime mould {\itshape Physarum polycephalum} is a suitable candidate organism for soft-matter robotics because it exhibits controllable transport, movement and guidance behaviour. \Phys may be considered as a smart computing and actuating material since both its motor and control systems are distributed within its undifferentiated tissue and can survive trauma such as excision, fission and fusion of plasmodia. Thus it may be suitable for exploring the generation and distribution of micro-actuation in individual units or planar arrays. We experimentally show how the plasmodium of \Phys  is shaped to execute controllable oscillatory transport behaviour applicable in small hybrid engines. We measure the lifting force of the plasmodium and demonstrate how protoplasmic transport can be influenced by externally applied illumination stimuli. We provide an exemplar vehicle mechanism by coupling the oscillations of the plasmodium to drive the wheels of a Braitenberg vehicle and use light stimuli to effect a steering mechanism. Using a particle model of \Phys we show how emergent travelling wave patterns produced by competing oscillatory domains may be used to to generate spatially represented actuation patterns. We demonstrate different patterns of controllable motion, including linear, reciprocal, rotational and helical, and demonstrate in simulation how dynamic oscillatory patterns may be translated into motive forces for simple transport of substances within a patterned environment.
\end{abstract}


\section{Introduction}\label{sec:introduction}

Very small scale mechanisms made possible by recent advances in micro-fabrication still require a means of energy conversion to convert a fuel supply into useful motive force. Although it is possible to reduce conventional engines and transmission designs in scale, doing so introduces challenges with regard to friction~\cite{fearing2002survey}, heat~\cite{ai2005heat} and more general miniaturisation problems~\cite{sher2009miniaturization} that may differ from or exacerbate those observed at the larger scale. 
Recent approaches have sought inspiration from biological sources of to provide the transformation of energy conversion to motive force, examples from flagellated movement of bacteria~\cite{zhang2009artificial} and eukaryotes~\cite{dreyfus2005microscopic}, ciliated transport for microscopic mixing~\cite{den2008artificial} and transport~\cite{suh2002fully}, peristaltic propulsion~\cite{trimmer2006caterpillar, saga2004development}, amoeboid movement~\cite{umedachi2009modular} and collective mass transport~\cite{kubea2000cooperative} have all been suggested. 

Biologically inspired mechanisms are attractive since they are composed of simple and plentiful materials and show impressive feats of redundancy, fault tolerance and self repair. However, even though the above mechanisms are technologically elegant and the product of impressive biological self-assembly, the internal arrangement of components underlying some mechanisms --- for example the bacterial flagellar motor system --- are as complex as those in non-biological machines. Progress in the development of biologically inspired machines has also been made by considering the use of even more simple biological structures which straddle the boundary of non-living physical materials and living organisms. Such structures include biological fibres and membranes~\cite{zhang2003fabrication}, lipid self assembly in terms of networks~\cite{lobovkina2008shape}, pseudopodium-like membrane extension~\cite{lobovkina2010protrusivegrowth} and even basic chemotaxis response~\cite{lagzichemodroplets2010}. 

Of particular interest are materials which embody oscillatory states, since the periodic transitions between two or more states can provide differential behaviour and impetus, if suitably coupled to a propulsion mechanism. The resulting transport may be relatively simple, such as flagellar or ciliary motion~\cite{dillon2007fluid}, or as complex as the patterns produced by neural assemblies of central pattern generators~\cite{ijspeert2008central}. When oscillatory states are combined with local communication waves of excitation may propagate which can also result in motion if the excitation is coupled to propulsion mechanism. Reaction-diffusion phenomena have proven to be useful in this approach because the emergence of oscillatory states can be a self organised and robust process and also feature travelling wave phenomena. Reaction-diffusion approaches to robotics include the use of propagating waves for robotic control and navigation~\cite{adamatzky2004rdrobot}, for the manipulation and transport of objects~\cite{adamtazky2006manipulating}, and oscillatory wave transport within a gel-like material for forward movement~\cite{maeda2007self}.

Biological (complete living systems) and semi-biological (materials produced by living systems) approaches can each provide valuable insights and techniques in the pursuit of small scale machines. The ideal hypothetical candidate for a biological machine would be an organism which is capable of the complex sensory integration, movement and adaptation of a living organism, yet which is also composed of a relatively simple material that is amenable to simple understanding and control of its properties. 

We suggest that the myxomycete organism, the true slime mould {\itshape Physarum polycephalum}, is a suitable candidate organism which meets both criteria; i.e. it is a complex organism, but which is composed of relatively simple materials. A giant single-celled organism, \Phys  is an attractive biological candidate medium for emergent motive force because the basic physical mechanism during the plasmodium stage of its life cycle is a self-organised system of oscillatory contractile activity which is used in the pumping and distribution of nutrients within its internal transport network. The organism is remarkable in that the control of the oscillatory behaviour is distributed throughout the almost homogeneous medium and is highly redundant, having no critical or unique components. 

The plasmodium of \Phys  is amorphous, jelly-like and flat in appearance and ranges from the microscopic scale to up to many square metres in size. The plasmodium is a single cell formed by repeated nuclear division and is comprised of a sponge-like actin-myosin mesh-work presented in two physical phases. The gel phase is a dense matrix subject to spontaneous contraction and relaxation under the influence of changing concentrations of intracellular chemicals. The sol phase is a liquid protoplasm transported through the plasmodium by the force generated by the oscillatory contractions within the gel matrix. There is a complex interplay between the gel and sol phases and both phases can change between each form when subject to changes in pressure, temperature, humidity and local transport. The internal structure of \Phys  can thus be regarded as a complex functional material capable of both sensory and motor behaviour. Indeed \Phys  has been described as a membrane bound reaction-diffusion system~\cite{adamatzkynaturwissenschaften} in reference to both the complex interactions within the plasmodium and the rich computational properties afforded by its 'material' properties \cite{adamatzky2008bzphysarum,AdamatzkyPhysarumMachines}. 

From a robotics perspective it has previously been shown that plasmodium of \Phys, by its adaptation to changing conditions within its environment, may be considered as a prototype micro-mechanical manipulation system, capable of simple and programmable robotic actions including the manipulation (pushing and pulling) of small scale objects~\cite{adamatzky2008towards,AdamatzkyPhysarumMachines}, light-controlled 
propelling of small floating objects~\cite{adamatzkyPhysarumBoats, AdamatzkyPhysarumMachines},
peristaltic transporter (intelligent self-growing pipe) of biologically friendly substances~\cite{AdamatzkyPhysarumMachines,AdamatzkyPainting},
 and as a guidance mechanism in a biological/mechanical hybrid approach where the response of the plasmodium to light irradiation was used to provide feedback control to a robotic system~\cite{TsudaS05RobotCtrlCellPreProc}. A \Phys  inspired approach to robotics using amoeboid movement generated by a series of fluid coupled oscillators has been also been demonstrated~\cite{umedachi2009modular}.

In this report we discuss the use of \Phys  as a candidate organism and material for the spontaneous generation and distribution of physical forces for engine-like and distributed planar transport mechanisms. The layout of the paper is as follows. 
Section~\ref{sec:experimental} shows experimental results demonstrating that \Phys  plasmodium is capable of generating motive lifting force which is both measurable and externally controllable. Using a dumbbell patterned design we demonstrate  complex phase relationships from the oscillatory dynamics and we design a prototype example mechanism where the motive output of \Phys  plasmodium oscillations is used to drive the wheels of a simulated Braitenberg vehicle, using external light stimuli to steer the vehicle. Building upon the experimental results we investigate in Sect.~\ref{sec:model} the emergence of oscillatory transport in the computational modelling of a virtual material composed of a particle collective where --- as with the plasmodium --- oscillatory phenomena emerge from simple and local interactions. Building on the so-called `smart table' approach in ~\cite{adamtazky2006manipulating}, we suggest a \Phys inspired method of generating wave based transport by patterning a planar environment into predefined shapes. In Sect.~\ref{sec:results} we demonstrate a range of transport types (linear, rotary, reciprocal and helical) and coupling methods using this approach. We also demonstrate in simulation how the emergent oscillatory patterns in the model may be interpreted as a vector field to transport material within patterned regions. We summarise the experimental and theoretical findings in Sect.~\ref{sec:discussion} and suggest future research by which small scale transport devices may be constructed from materials which exhibit emergent properties from simple component interactions.

\section{Experimental}\label{sec:experimental}

This section focus on experimental investigation into the possibility of
\Phys  engine implementation. The \Phys 
plasmodium is famously known for fast protoplasmic streaming within the
cell body and the motive force of the cell has been studied extensively
in the field of cell biology for more than 50 years
(cf.~\cite{allenpj50physarumrespirationandprotoflow,KamiyaN50ProtoFlow,GotohK82MotiveForcePhysarum}). The
mechanism of streaming generation is based on the actin
polymerization/depolymerization cycle, which switches every 30~s
regulated by calcium ions~\cite{SmithDA92CaModelPhysarum}. This periodic
cycle controls the direction of the protoplasmic streaming, and as a
result, the whole cell shows rhythmic cell thickness changes~\cite{KamiyaN50ProtoFlow}. 
The flow speed of protoplasmic streaming of
the plasmodium becomes up to 1~mm/s, whereas that of
other organisms, such as plant and amoeba cells, are about tens of
$\mu$m/s~\cite{KamiyaN58VelcDistPhysarum}. This is one of the fastest
protoplasmic streaming known so far, and together with the periodicity
of the protoplasmic streaming, the plasmodium of \Phys could be ideal 
candidate to fabricate small-scale biological devices, e.g. actuator and transporter.

In order to explore the possibility of using the \Phys  cell
as source of power, it is important to know (a)~how much force a
\Phys  plasmodium can generate, and (b)~how the force can be
steered externally. Before going to these two points, we first discuss
characteristics of the cell shape and the contraction oscillation rhythm
in the \Phys  plasmodium.

\subsection{Cell Shape and Oscillation Pattern}

\begin{figure}[tbp!]
 \centering
 \subfigure[]{\includegraphics[width=0.38\textwidth]{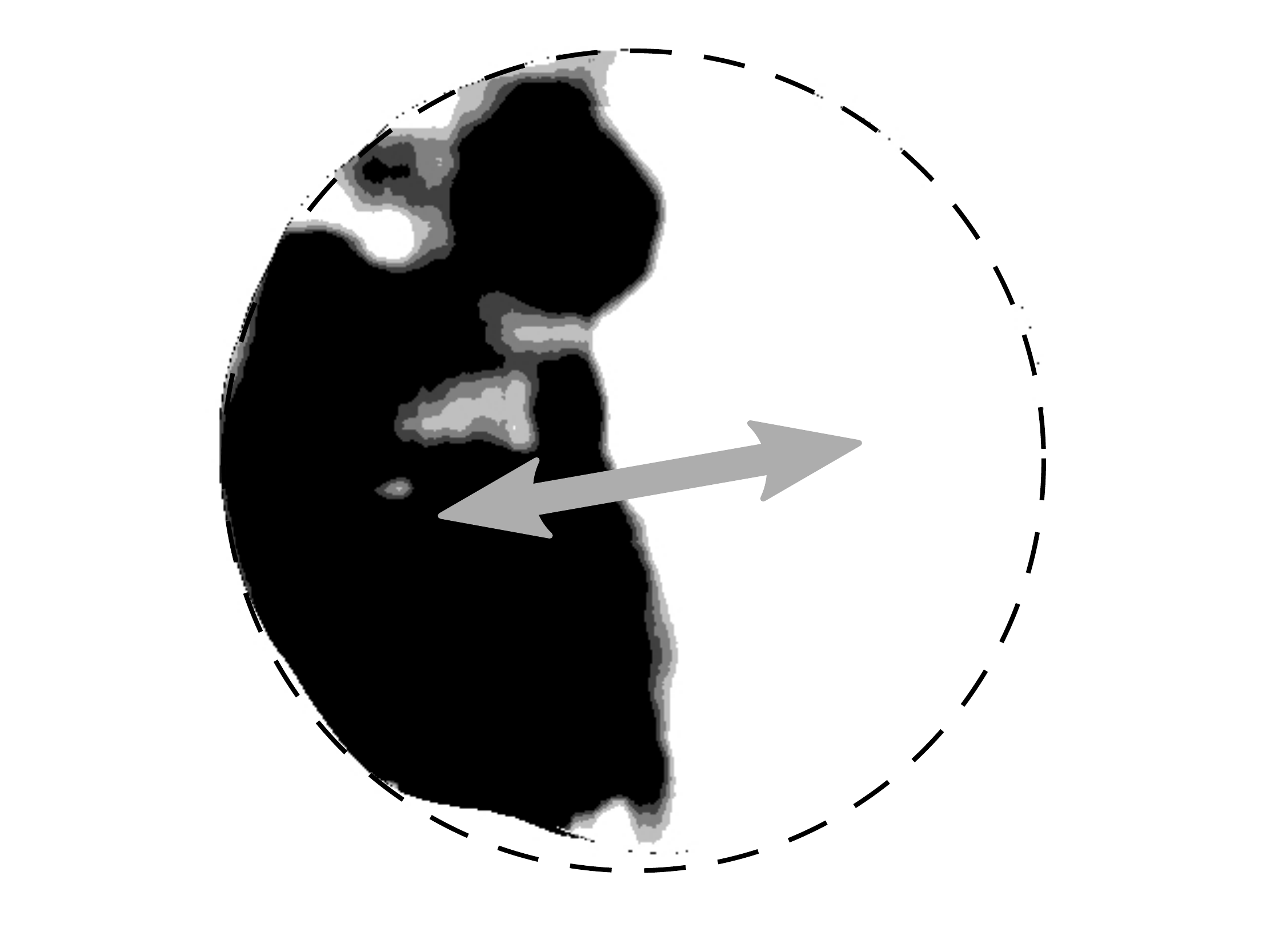}}
 \subfigure[]{\includegraphics[width=0.38\textwidth]{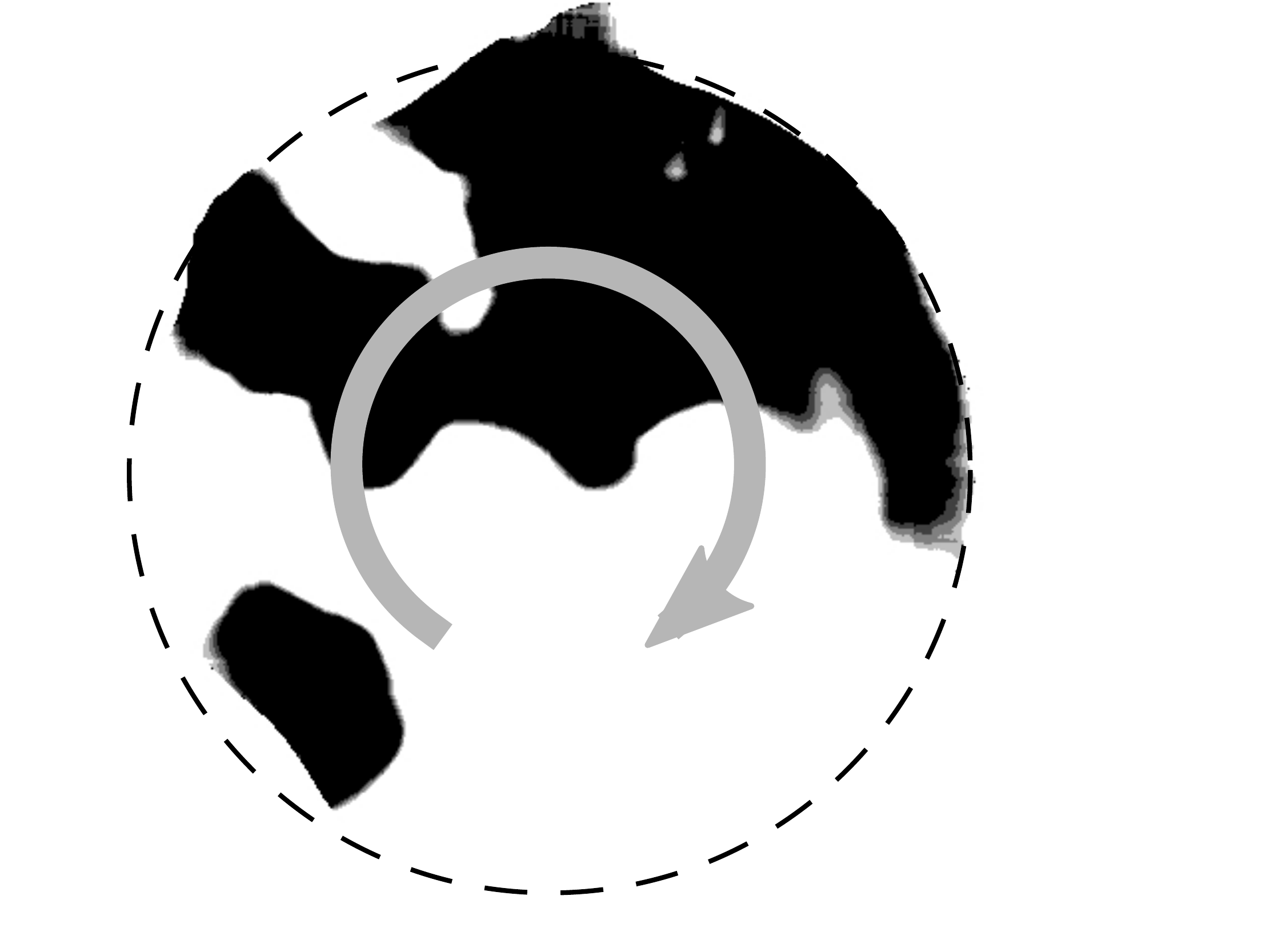}}
  \caption{Oscillation patterns of the \Phys  plasmodium in
 a single well: (a)~bilateral shuttle oscillation and (b)~clockwise rotation
 wave. Black and white regions indicate areas where thickness is
 increasing and decreasing, respectively.}  \label{fig:phyoscipatterns}
\end{figure}

The generation of periodic rhythm of the cell confined in a small
circular well on the agar gel was investigated by the authors and also
other researchers
previously~\cite{TsudaS10PhyOsciBiosystems,TakagiS08EmergentOsciPatternPhysarum}.  
When observed under a microscope,
a piece of \Phys  cell just placed in the well consists of
smaller pieces of granular protoplasm, each of which can potentially
become an individual slime mold cell. These granules start contracting
about 10~min after transplanted and neighboring granules gradually
merge together to synchronize the contraction rhythm. They eventually
fuse into one single \Phys  plasmodium and show several
types of synchronized oscillation rhythms, such as bilateral shuttle
streaming of protoplasm (Fig.~\ref{fig:phyoscipatterns}a) and
clockwise or anti-clockwise rotation streaming
(Fig.~\ref{fig:phyoscipatterns}b). What is remarkable about the
oscillation regeneration in the \Phys  plasmodium is
``size-invariant'' behavior. Even if the cell size becomes larger with 
increased size of well, a whole cell fully synchronizes relatively quickly. 
We undertook experiments with 1.5, 3.0, 4.5, 6.0, and 7.5~mm diameter wells, 
and in all cases the \Phys  cell fully synchronized approximately in 70~min~\cite{TsudaS10PhyOsciBiosystems}. 
Thus the plasmodium is good at coordinating the internal structure while maintaining
itself as a single cell. Even when the body size changes, the plasmodium restores
the synchronization using a distributed-computing type control.

Although a cell in the well does show oscillations that potentially can
be exploited as source of power, the single well design would not be
appropriate for \Phys  engine. Given that a whole plasmodium in a well is used to apply force
to another object, a total displacement (i.e. thickness change) of the
cell will be crucial. However, in the case of single well, the total
displacement will be balanced out. For example, if a cell is showing
bilateral oscillation (Fig.~\ref{fig:phyoscipatterns}a), the thickness
of a half of the cell is increasing while that of the other half is
decreasing. As a result, the total displacement becomes almost
zero. This fact led us to the design of a dumb-bell shape, in which two
circular wells are connected by a narrow channel, see Fig.~\ref{fig:steeringphyoscillation}a. 
This design is originally
made by Takamatsu et al~\cite{TakamatsuA02ConstLivCoupOsciPhysarum}.
When the width of the connecting channel is 0.4~mm, anti-phase thickness
oscillation between two wells can be observed~\cite{TakamatsuA00CtrlIntStrnPhsarumOsci}. 
This means protoplasm of the cell flows back and forth between wells, and thus the 
thickness displacement in one well will not be canceled out.

Therefore we adopt the dumb-bell shape design and keep
the cell to take the shape by hydrophobic structures. A plasmodium prefers
high-humidity niche and it stays inside the structures of wet-wells and
keeps its shape for a long time. \Phys  cells used for
experiments below were cultured on 1.5~\% agar gel in the thermostatic
chamber (Lucky Reptile Herp Nursery II Incubator, Net Pet Shop, UK) at
$26^{\circ}$C and fed with oat flakes once or twice a day. They were
starved at least for 12~hours prior to experiments.

\subsection{Force generated by plasmodium}\label{sec:force-gener-itsh}

\begin{figure}[tbp!]
 \centering
 \subfigure[]{\includegraphics[width=0.45\textwidth]{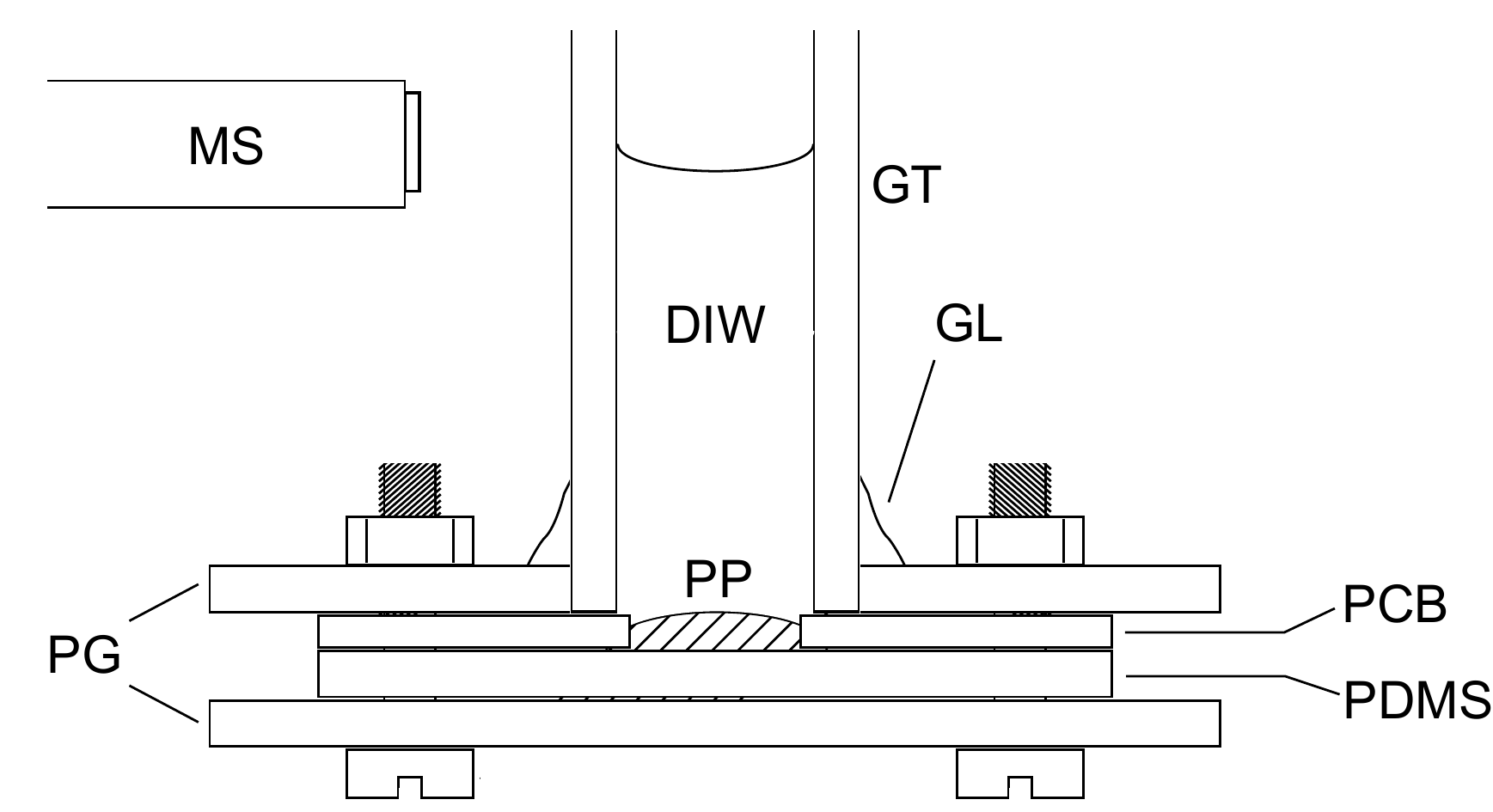}}
 \subfigure[]{\includegraphics[width=0.4\textwidth]{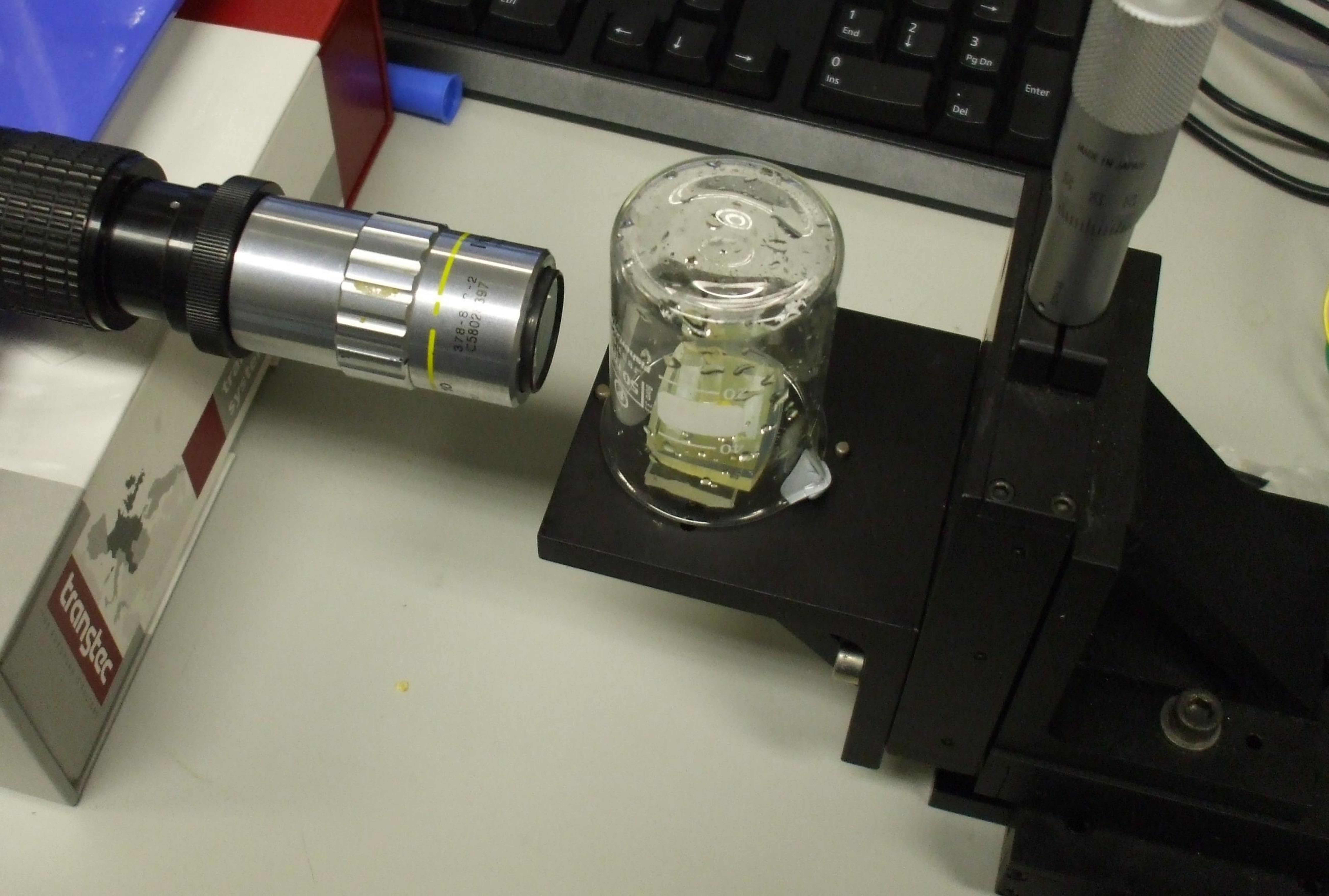}}
 \caption{Experimental setup for plasmodium force measurement:
 (a)~A schematic illustration of plamodial force measurement
 setup. MS: Microscope, GT: Glass tube, DIW: Distilled
 water, GL: Glue, PP: \Phys  plasmodium, PG: Plexi glass,
 PCB: Printed circuit board, PDMS: Polydimethylsiloxane silicon
 rubber. (b)~A photo of the setup. }
 \label{fig:pressuresetup}
\end{figure}

Let us evaluate how much force could be generated plasmodium's  oscillatory
contractions. We estimate the force by loading some weight on a plasmodium:
we load water one of two wells of a dumb-bell shaped plasmodium (Fig.~\ref{fig:pressuresetup}a). 
By changing the height of water, the load on the cell is controlled. 

A dumb-bell shaped plasmodium was prepared as follows. Small
pieces of \Phys  are taken from a larger culture and set in
dumb-bell shaped wells of a thin printed circuit board (PCB), with 
copper coating removed. They are then placed on a 1.5~\% agar gel 
and kept in a dark place at least for two hours so that the pieces of 
\Phys fuse into one single cell. After the fusion occurs, the cell 
and the PCB are clamped together with plexiglass and a PDMS block
(Fig.~\ref{fig:pressuresetup}a). A glass tube, which inner diameter is
same as the diameter of a plasmodium well, is glued to the top plexi
glass and filled with distilled water to a certain height. The whole
setup is covered with a wet glass beaker to avoid evaporation of water
in the glass tube and placed on the three-dimensional micro-stage
(Fig.~\ref{fig:pressuresetup}b). The displacement of water surface is
recoded for 20~min by a microscope from the size. 

\begin{figure}[tbp!]
\centering
\subfigure[]{\includegraphics[width=0.6\textwidth]{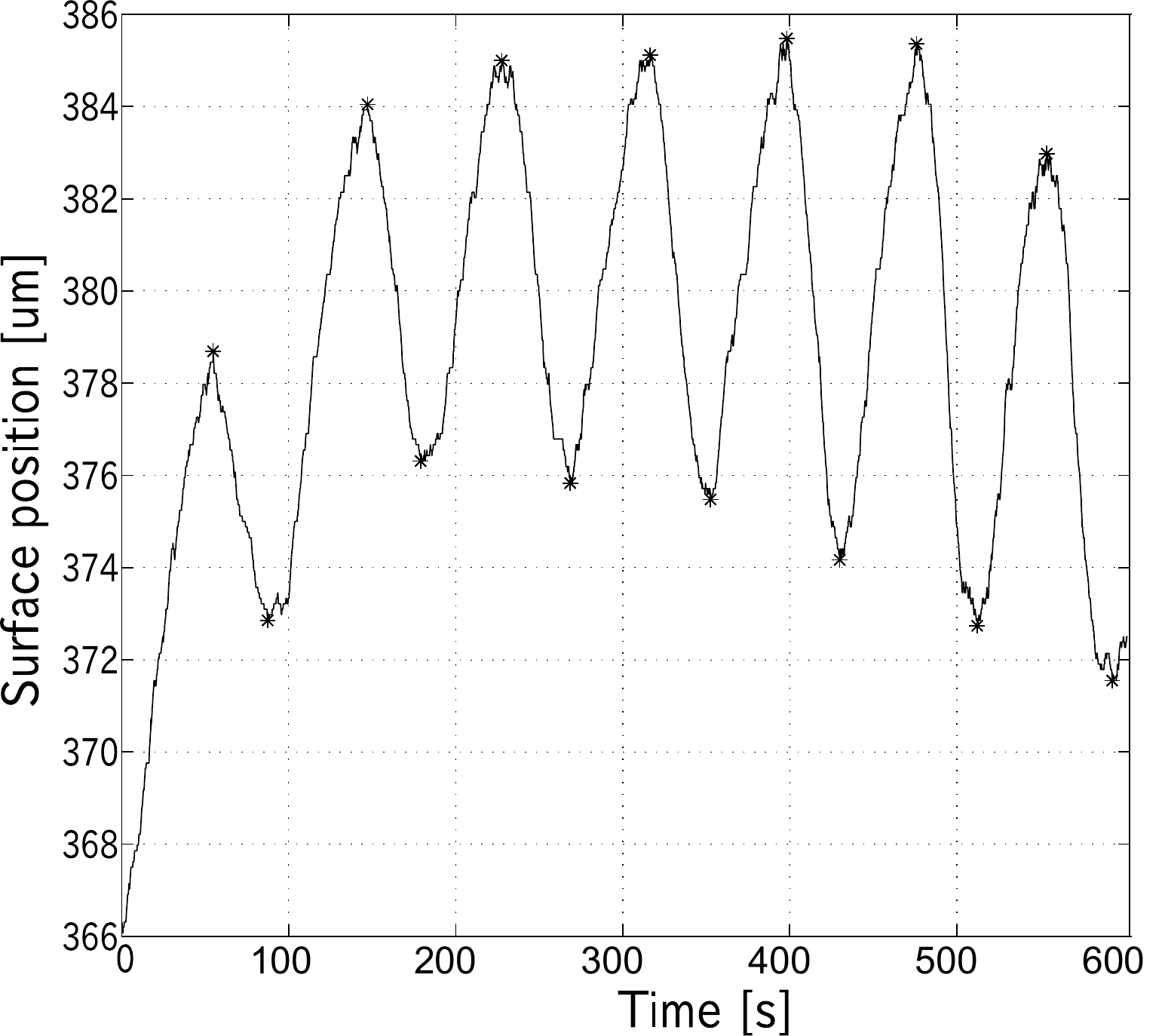}}
\subfigure[]{
   \begin{tabular}[t]{|c|c|}
   \hline
   Water height & Max. displacement \\
   \hline
   40~mm &  20~$\mu$m \\
   60~mm &  14~$\mu$m \\
   80~mm &  10~$\mu$m \\
   100~mm & 7~$\mu$m \\
   \hline
   \end{tabular}
  }
  \caption{(a)~A time course of the surface displacement in the case of
 100~mm water height. First 10~min is shown. (b)~A table of the maximum
 water displacement with different water height. It decreases
 monotonically as the water height increases.}
 \label{fig:surfacedisplacement}
\end{figure}

Figure~\ref{fig:surfacedisplacement}a shows a result of surface
displacement experiment in the case of 100~mm water height. We have
tested with 40, 60, 80, and 100~mm water height (71, 106, 141, and
177~mg in weight, respectively). 

In all the cases the height of water surface periodically oscillates due
to the thickness oscillation of the plasmodium at the bottom.  When no
load is placed on the slime mould, it showed thickness oscillations of
approximately 60s per cycle, which is a typical oscillation period of
the {\itshape Physarum} plasmodium. When DIW weight was added on the
cell, on the other hand, the period became longer up to twofold. The
more load the slime mould carries, the slower the period became. For
example, a typical period was 65s and 90s per cycle for 20mm and
60mm-high water weight, respectively. Even at the maximum load,
100mm-high case, it was around 100s per cycle, as shown in
Fig.\ref{fig:surfacedisplacement}a. During an experiment with a fixed
load, the period slowed down gradually and slightly. In many cases,
water height slightly decreased (approximately 10--40 $\mu$m) during a
20~min experiment. This is possibly because the \Phys cell partially
escaped to the side of the well not connected to the water tube. The
maximum displacement of water surface was estimated by detecting peaks
and bottoms of the oscillation (marked as ``$\ast$'' in
Figure~\ref{fig:surfacedisplacement}a) and calculated differences from a
bottom to the next peak. Figure~\ref{fig:surfacedisplacement}b
summarizes the maximum displacement of various water heights. As the
water height becomes higher, the displacement decreases monotonically.
By extrapolating this result, the maximum height the dumb-bell shaped
plasmodium could bear with would be 140~mm, and it could lift up to
approximately 250mg load.

The work by the plasmodium in the case of 100~mm water height can be
calculated as $W = (0.75)^2\times \pi\times 100\times \rho\times g\times
0.007 \approx 12.1 \ \ \textrm{(mJ)}$, where $\rho$ is the density of
water (1mg/mm$^3$) and $g$ is the gravity acceleration
($9.8$~m/s$^2$). Assuming that the speed of thickness change is
constant, the power of the cell can be estimated as $ P = W/T = 12.1 /
50 \approx 0.24 \ \ \textrm{(mW)}$.  

At the micron scale, however, one has to bear in mind that some factors,
which do not need to be considered at a macro scale, become crucial. For
example, friction becomes relatively large compared to the force
generated by the cell. Thus, for the effective use of the force, a
careful attention to things like a choice of lubricant and the design of
stop valve have to be paid. Nevertheless, it is noteworthy that a tiny
cell, which weight approximately 5~mg, can lift up a load over 36 times
heavier than its own weight. Such strong force generated by the
{\itshape Physarum} plasmodium was in fact reported more than three
decades ago by several
researchers~\cite{YoshimotoY78Contractrhythm,Wohlfarth-Bottermann_1977}.

\subsection{Steering Control of \Phys  Engine}\label{sec:steer-contr-itsh}

There are several ways to externally stimulate the \Phys's
oscillation, such as light~\cite{HaederDP85Ca2PhototaxisPhysarum},
temperature~\cite{WolfR97PhysTempGrad}, and electric stimuli~\cite{AndersonJD51GalvanotaxisPhysarum}. 
Among others, light stimulus is
the most commonly used one because it can be easily introduced or
removed simply by switching light on and off. It is known that, when a
local part of a plasmodium is exposed to white light, the frequency of
thickness oscillation at the local part decreases
\cite{NakagakiT99ModCellRhythm}.  Hence we adopt white light as
external controller for steering of \Phys  engine.

\begin{figure}[tbp!]
 \centering
 \subfigure[]{\includegraphics[width=0.4\textwidth]{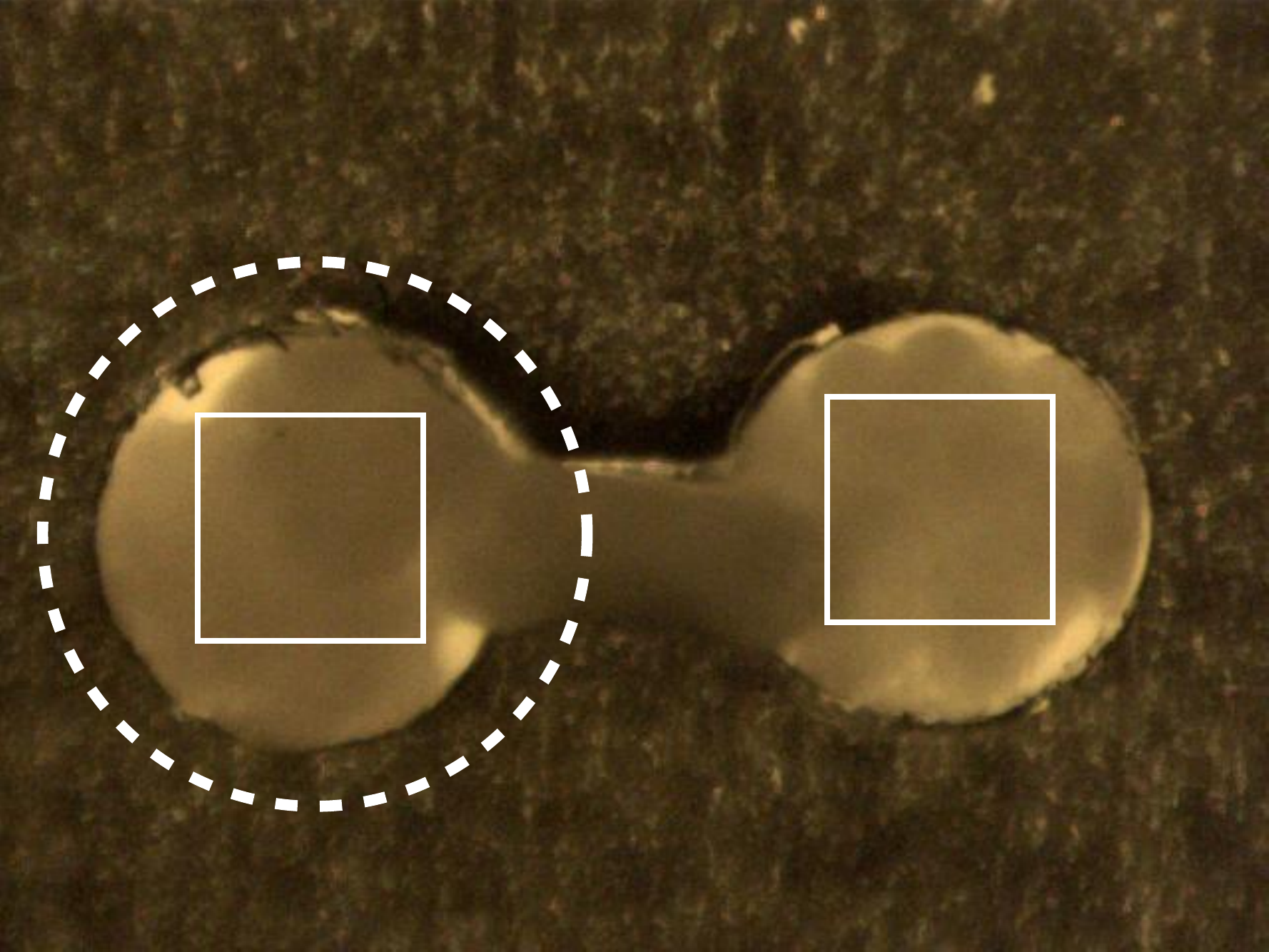}}
 \subfigure[]{\includegraphics[width=0.5\textwidth]{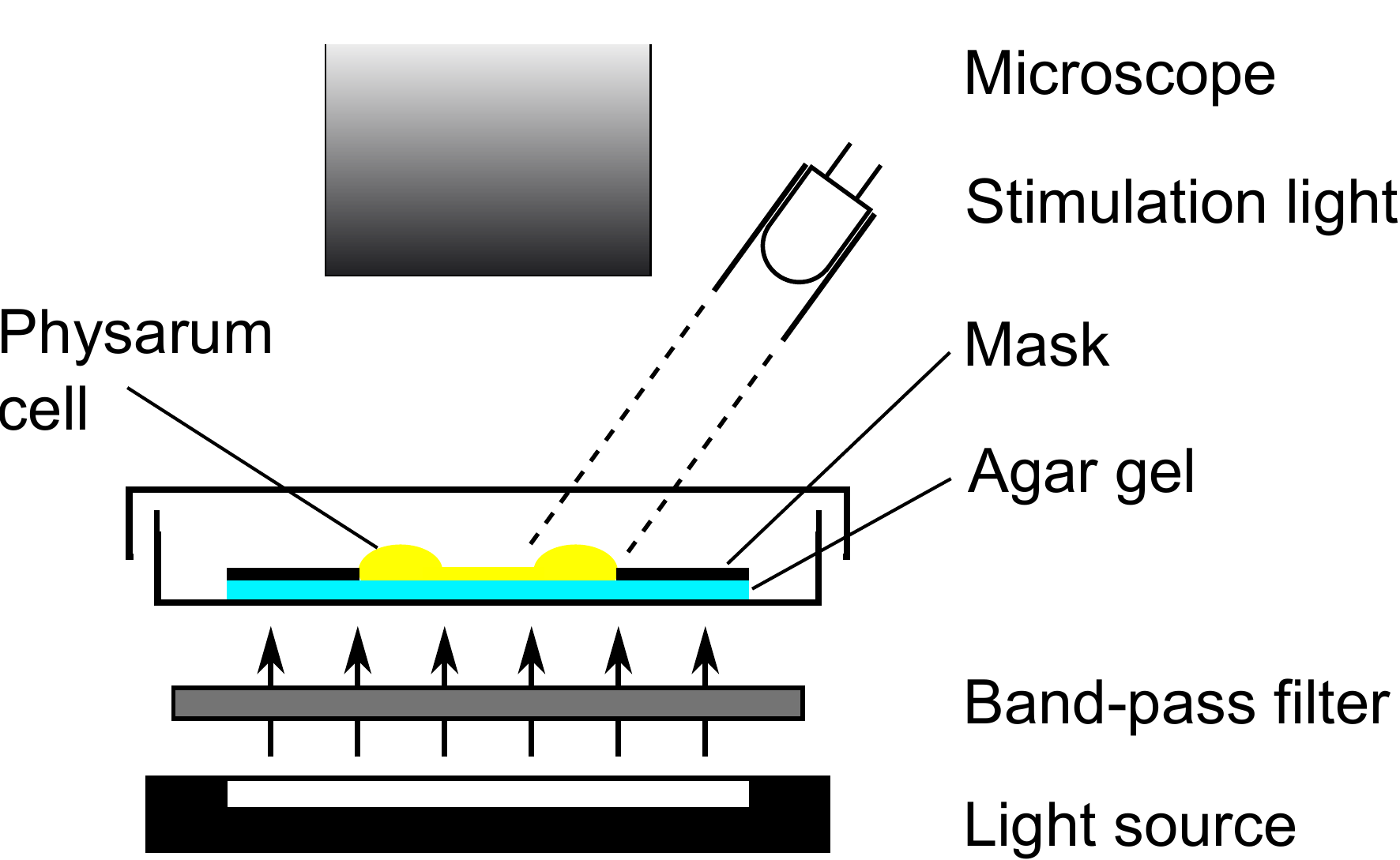}}
 \caption{(a)~A dumb-bell shaped plasmodium on a 1.5\% agar gel. The
 black mask surrounding the cell is made of transparency sheet. A dotted
 circle indicates the area exposed to white light. The areas indicated
 by two solid squares are used to calculate thickness oscillations of
 the cell.  (b)~A schematic illustration of light illumination
 experiment. The cell is exposed to bandpass-filtered light from the
 bottom for monitoring and one of two wells is exposed to white LED
 light from the top for stimulation. The cell is kept in a Petri dish to
 maintain constant humidity condition. }
 \label{fig:steeringphyoscillation}
\end{figure}

To confirm if white light affects the oscillation rhythm of the
dumb-bell shaped cell, we conducted the following experiment. The
dumb-bell shaped plasmodium is constructed on on a
1.5~\% agar gel using a transparency sheet mask
(Fig.~\ref{fig:steeringphyoscillation}a). The dumb-bell shaped hole in
the mask was cut out with a 1/16'' hand punch, which makes approximately
1.5~mm diameter hole, and the connecting channel in was cut with
a scalpel under a stereo-microscope. The structure is placed under a microscope
(Leica Zoom 2000, Germany) and illuminated from the bottom with
bandpass-filtered light near the 600~nm wavelength
(Fig.~\ref{fig:steeringphyoscillation}b). This wavelength of light does
not affect the \Phys 's oscillation activity
\cite{NakagakiT96ActSpectSporPhoto}. The thickness of the cell is
measured by the light transmission through the cell, which is inversely
proportional to the thickness. The controller, an ultra bright white LED, 
illuminates only one of two wells of
the plasmodium (a dotted circle in
Fig.~\ref{fig:steeringphyoscillation}a). A snapshot image of the cell is
saved every 3~s in a PC for subsequent image analysis.

The thickness of the cell is calculated as follows. We convert snapshots
from RGB to gray scale, and then calculate a difference of images taken 
at $t$ and $t-\Delta t_1$. This gives us a thickness change of the 
plasmodium in $\Delta t$ (we used $\Delta t_1=7$). Then we
apply a moving average filter spatially over 91$\times$91 pixels 
(indicated by a solid square
in Fig.~\ref{fig:steeringphyoscillation}a) and temporally over 15 images
on each well. This works as an image smoothing filter to reduce camera
noise.

\medskip

\begin{figure}
 \begin{center}
  \includegraphics[width=0.75\textwidth]{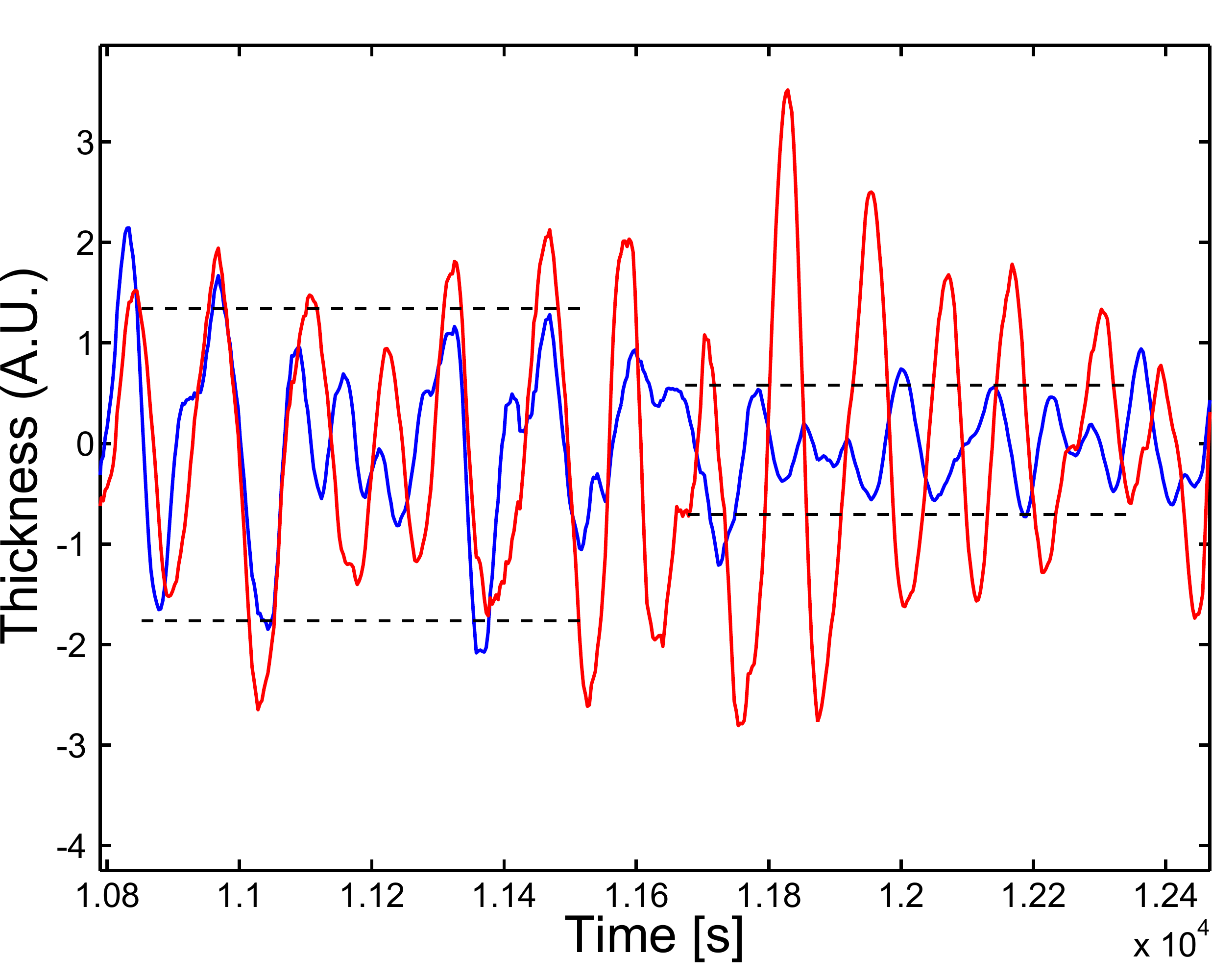}
 \end{center}
 \caption{A typical reaction of the \Phys  plasmodium to
 light stimulus. Only one of two wells is exposed to white light (blue
 curve). A LED light is turned on at the point indicated by a downward
 arrow and kept on till the end of the plot. The oscillation period of
 the stimulated well (blue curve) becomes slightly longer when the light
 is on, whereas that of the unstimulated well (red curve) remains unchanged.
 The oscillation amplitude significantly decreases in the stimulated well
 and increases in the other well (horizontal lines).} 
 \label{fig:steeringphyoscillation-result}
\end{figure}

Figure~\ref{fig:steeringphyoscillation-result} shows a typical reaction
of the cell to white light. Before the light is switched on the 
oscillations in two wells are synchronized in amplitude and phase
(the first half of Fig.~\ref{fig:steeringphyoscillation-result}). 
When one of the wells becomes exposed to the light, the oscillations
in wells go out of phase.  

Light is a negative stimulus to \Phys and slows down the contractile 
oscillation at the illuminated part of a plasmodium. Period of oscillations 
becomes longer in the latter half of the plot in Fig.~\ref{fig:steeringphyoscillation-result}, 
whereas those of unexposed well remain unchanged. 

Another reaction of the cell to the light is significant change in oscillation amplitudes. 
That in the stimulated well decreased largely soon after the light is turned on 
(indicated by horizontal dotted lines in Fig.~\ref{fig:steeringphyoscillation-result}), 
but on the other hand, that in the non-stimulated well increased.  This is because the 
cell is trying to escape from the exposed area by actively pumping the protoplasm from 
the stimulated well to non-stimulated well. When white light is removed, the oscillation at
the stimulated well came back to the original period and oscillation
amplitudes in both wells equalize.  Thus,
white light does slow down the dumb-bell shaped \Phys 
plasmodium, and with this light stimulus, the cell can be steered from
the outside. We are going to show a stimulated vehicle driven by the
\Phys  cell can be actually steered by light in the next
section.

\subsection{Vehicle simulation driven by experimental data}

\begin{figure}[tbp!]
 \centering
 \subfigure[]{\includegraphics[width=0.25\textwidth]{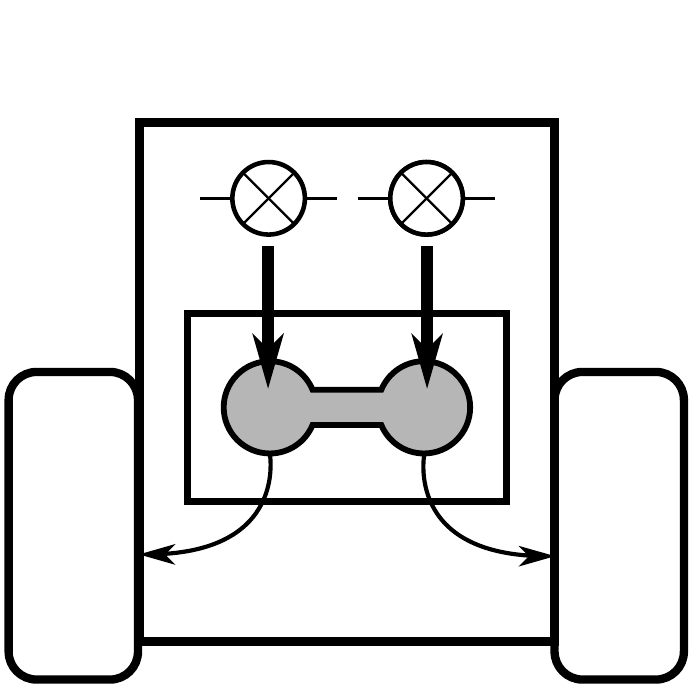}}
 \subfigure[]{\includegraphics[width=0.3\textwidth]{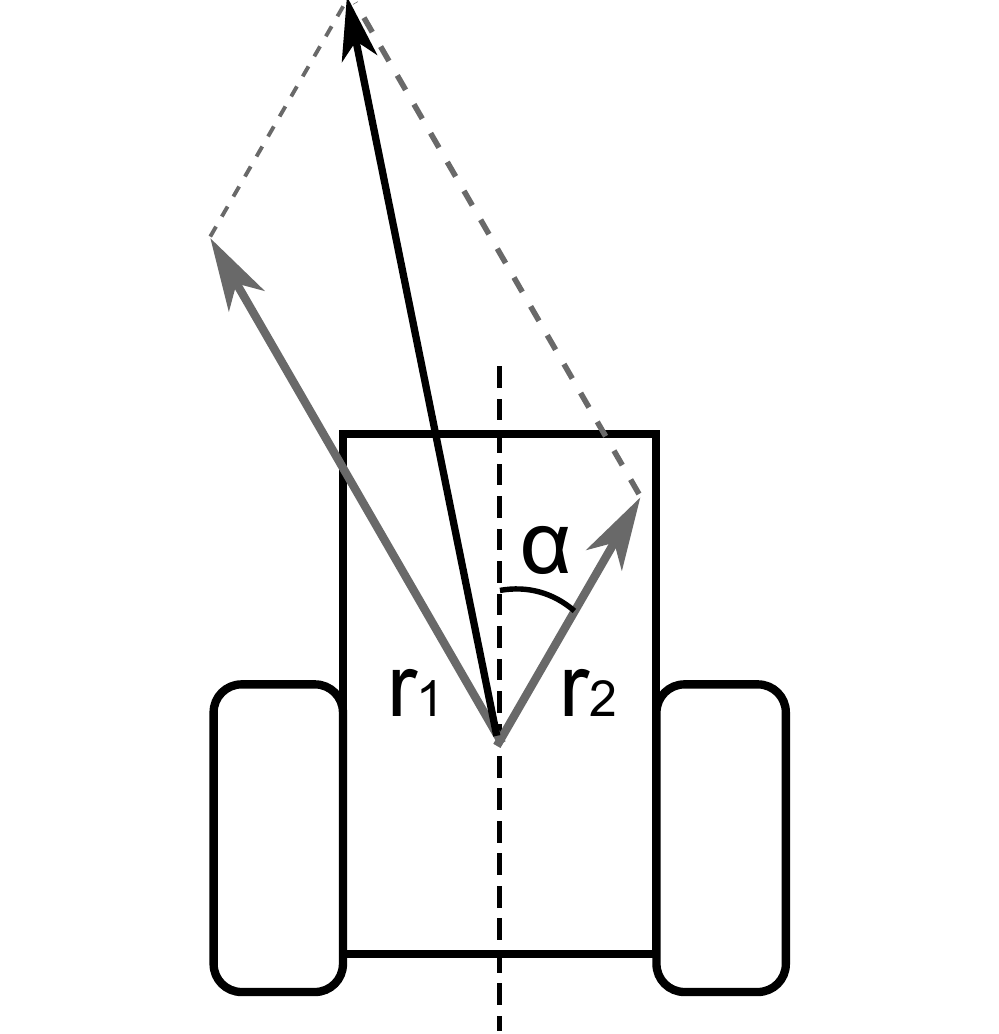}}
  \caption{(a)~A schematic diagram of a vehicle driven by {\itshape
 Physarum} engine. The oscillation amplitudes of the cell determines the
 wheel rotation speed of the vehicle. Light illumination onto each well
 of a {\itshape Phyasrum} plasmodium is used to steer the vehicle.  
 (b)~The motion of \Phys-driven simulated vehicle. Vehicle's
 direction of movement can be represented as addition of vectors
 (Eq.~\ref{eq:1}). }  \label{fig:vehicle}
\end{figure}

We assume that the power generated by a \Phys  plasmodium
is converted to drive wheels of a hypothetical micro-scale vehicle
without any energy loss. Increases and decreases in thickness
oscillations from two wells (Fig.~\ref{fig:steeringphyoscillation}a)
drive wheels of a Braitenberg-like vehicles
\cite{BraitenbergV84Vehicles} illustrated in Fig.~\ref{fig:vehicle}a.
The vehicle has two wheels driven by oscillations of a dumb-bell-shaped
plasmodium. It also has two lights which illuminate
two wells of the cell integrated in the vehicle in order to steer the
vehicle. 

There are several outputs of the thickness
oscillation that can be used to drive the wheels. For example, thickness
amplitude, oscillation frequency, and phase difference between
oscillations of two wells. Here we employ thickness amplitude to
demonstrate vehicle control. We assume that the thickness oscillation pushes a
hypothetical piston and rotates a crankshaft to change the speed of one
of wheels.  The wheels rotate at a constant speed and the thickness
oscillations of the cell increases the speed depending on the
amplitude. The motion of a \Phys-driven vehicle is illustrated in Fig.~\ref{fig:vehicle}b):

\begin{equation}\label{eq:1}
 \left\{\begin{array}{rcl}
  x(t+1) & = & x(t) +
   |r_1|\mathrm{sin}(\theta+\alpha)+|r_2|\mathrm{sin}(\theta-\alpha)\\ 
  y(t+1) & = & y(t) +
   |r_1|\mathrm{cos}(\theta+\alpha)+|r_2|\mathrm{cos}(\theta-\alpha)\\
	\end{array}\right.
\end{equation}
where $x(t)$ and $y(t)$ are positions of the vehicle at time $t$, and
$r_1$, $r_2$ are thickness displacement of the cell in left and right
wells in a certain period (in this case, 21~s). $\alpha$ is a parameter
that determines the contribution of the thickness displacement to the
change of wheel running speed. The larger $\alpha$ is, the more wheels
are accelerated. Hence, with larger $\alpha$, vehicle's direction of
movement becomes more sensitive to the difference of oscillation
amplitudes from two wells (i.e. if $||r_1|-|r_2||$ is large, the vehicle
turns either to the left or right).

\medskip

\begin{figure}[tbp!]
 \begin{center}
  \includegraphics[width=0.7\textwidth]{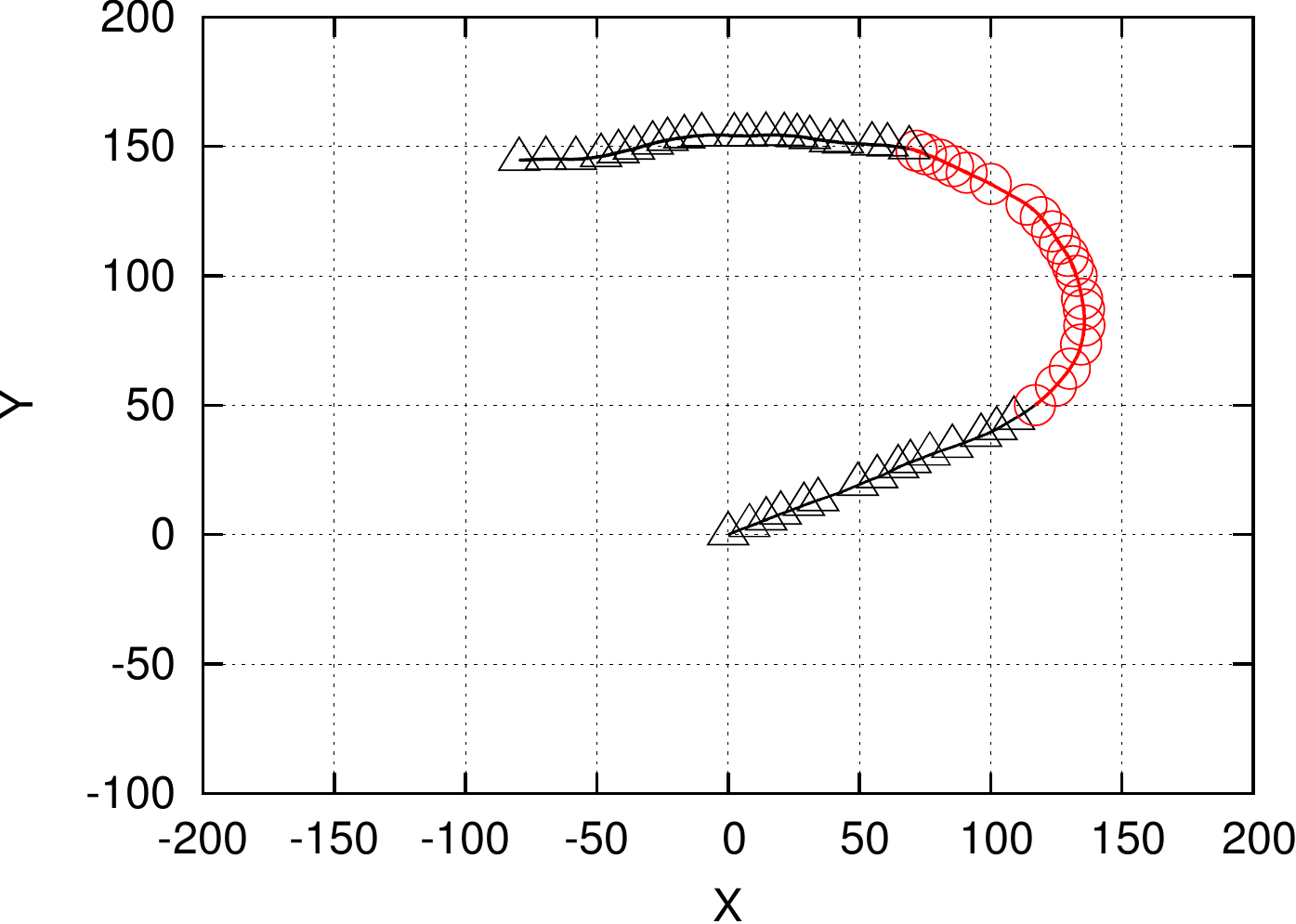}
 \end{center}
 \caption{Trajectory of the simulated vehicle driven by experimental
data. $\alpha = \pi/720$ is used in this simulation. The vehicle started
from (0,0). Black line indicates the period when there is no light
illumination to wells, and red line is when a right well is
illuminated by light.} \label{fig:vehicletraj}
\end{figure}

Figure~\ref{fig:vehicletraj} shows a trajectory of the simulated vehicle
driven by experimental data. The vehicle started from
the origin (0,0) goes straight (black line) when no light inputs are
on. It turns to the left (red line) as soon as the left well of the
dumb-bell-shaped cell is stimulated by light. When light is turned off,
it drives straight ahead again. This is can be explained as follows in
terms of \Phys's behavior. When there is no stimulus to
the cell, both wells oscillates with same amplitude
(cf. Fig.~\ref{fig:steeringphyoscillation-result}). This drives the
wheels with the same speed, and thus the vehicle goes straight. On the
other hand, as described in the previous section, the oscillation
amplitude in the stimulated well significantly decreases while the light
is on, whereas that in the other well increases. This is because the
cell moves out of the illuminated area and migrates towards the
non-illuminated well.  This slows down the left wheel driven by the
stimulated well and accelerates the right wheel. As a result, the
vehicle takes a left turn as long as the light is on. It drives in a
straight line again after the light is turned off because oscillation
amplitudes in two wells come back to nearly equal level when light
stimulation is removed.

\section{The Emergence of Oscillatory Transport Phenomena in a Particle-Based Model}\label{sec:model}

Due to the relatively slow time development of the \Phys  plasmodium and natural variability of its 'performance' in terms of the chosen robotics tasks (the plasmodium is only concerned with survival and not, after all, with satisfying externally applied experimental tasks), it is helpful to develop computational modelling approaches which may facilitate the study of distributed robotic control. Such a modelling approach allows us to explore the emergence of complex oscillatory behaviour from simple components and assess its application to robotics tasks.

\begin{figure}
 \begin{center}
  \includegraphics[width=0.95\textwidth]{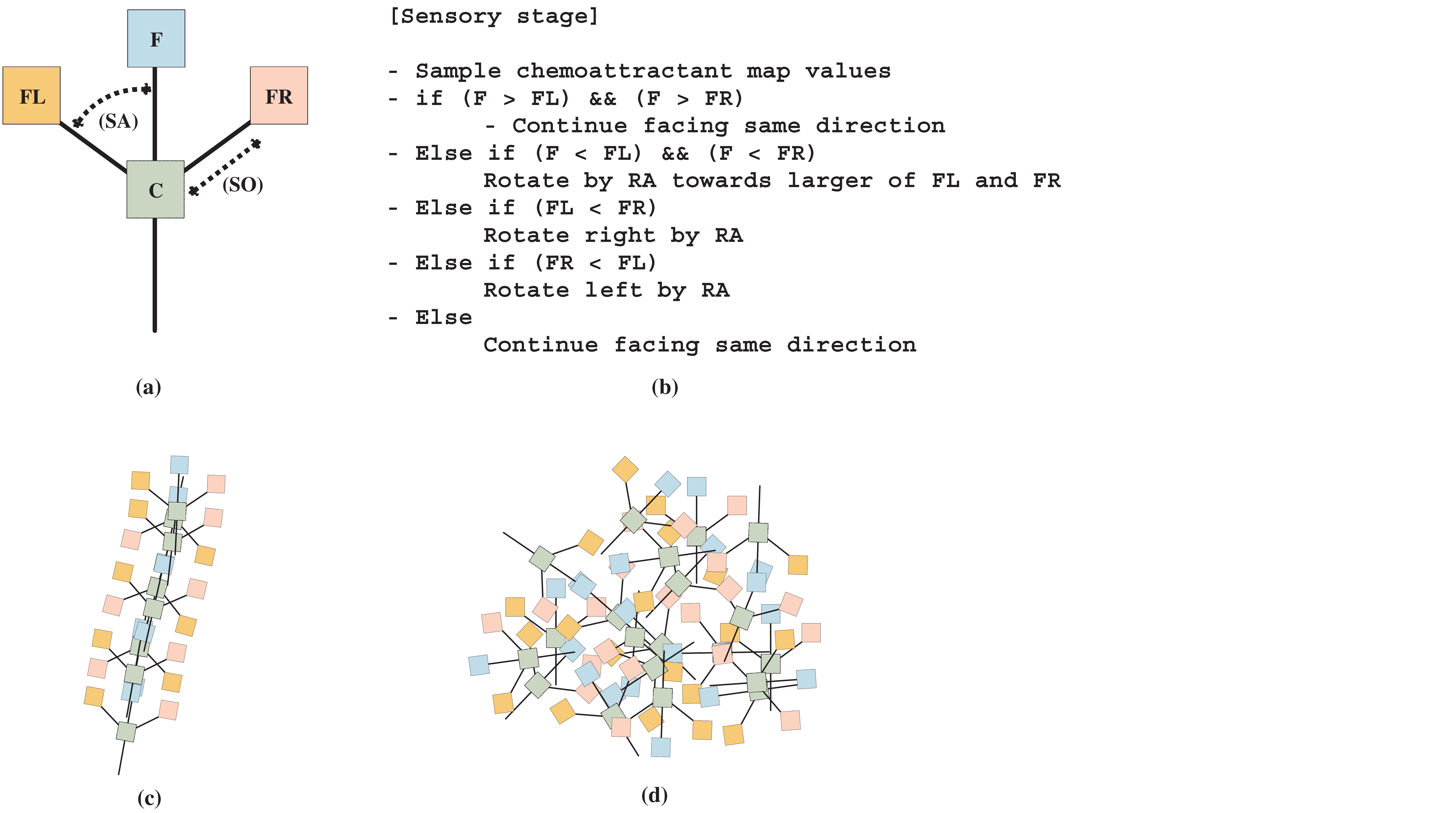}
 \end{center}
 \caption{Agent particle morphology, algorithm, and aggregation types.} 
 \label{particlemorphology}
\end{figure}

To investigate the use of emergent oscillatory phenomena for engine-like transport we employ an extension to the particle model in \cite{jones2010emergence} which was shown to generate dynamical emergent transport networks.  In this approach a plasmodium is composed of a population of mobile particles with very simple behaviours, residing within a 2D diffusive environment.  A discrete 2D lattice (where the features of the environment arena are mapped to greyscale values in a 2D image) stores particle positions and also the concentration of a local factor which we refer to generically as chemoattractant. The 'chemoattractant' factor actually represents the hypothetical flux of sol within the plasmodium.  Free particle movement represents the sol phase of the plasmodium. Particle positions represent the fixed gel structure (i.e. global pattern) of the plasmodium. Particles act independently and iteration of the particle population is performed randomly to avoid introducing any artifacts from sequential ordering. Particle behaviour is divided into two distinct stages, the sensory stage and the motor stage. In the sensory stage, the particles sample their local environment using three forward biased sensors whose angle from the forwards position (the sensor angle parameter, SA), and distance (sensor offset, SO) may be parametrically adjusted (Fig.~\ref{particlemorphology}a). The offset sensors represent the overlapping and inter-twining filaments within the transport networks and plasmodium, generating local coupling of sensory inputs and movement (Fig.~\ref{particlemorphology}c and d).

\begin{figure}[tbp!]
 \begin{center}
  \includegraphics[width=0.85\textwidth]{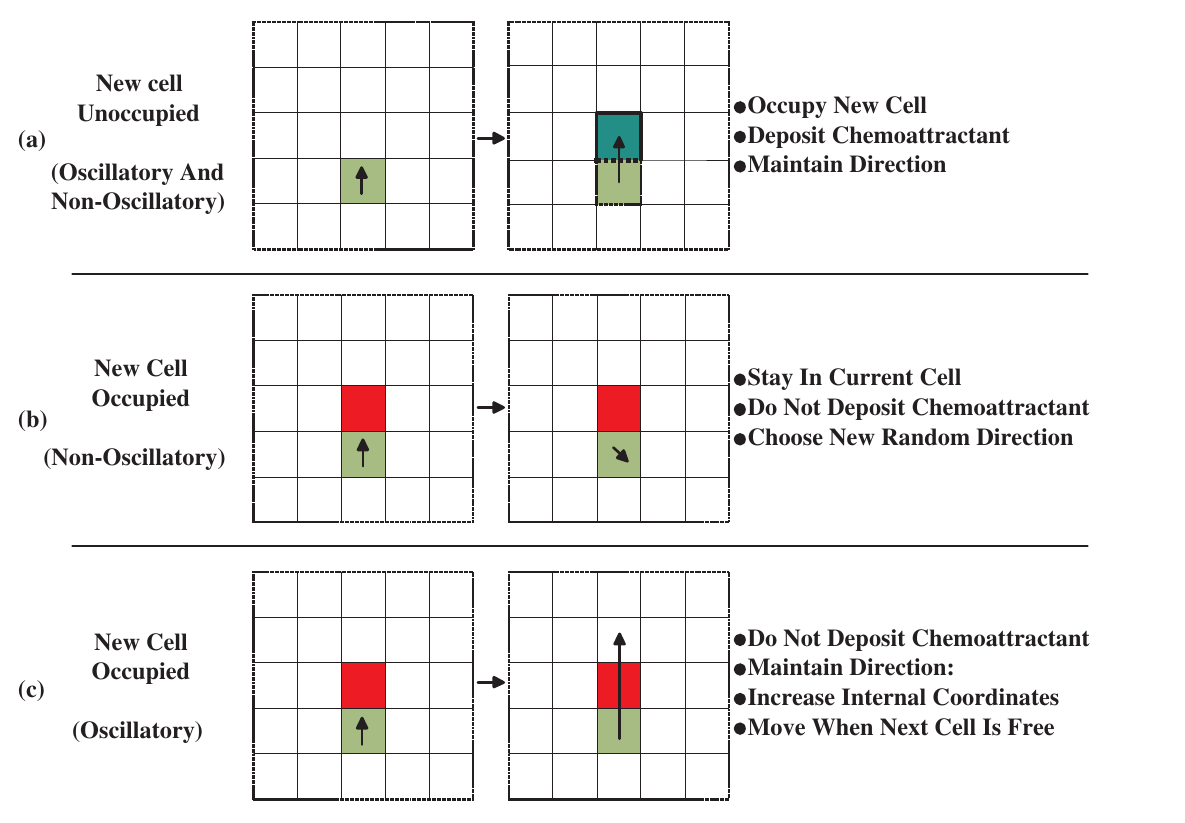}
 \end{center}
 \caption{Particle motor behaviour in non-oscillatory and oscillatory conditions} 
 \label{particlemotorbehaviour}
\end{figure}

The SO distance is measured in pixels and a minimum distance of 3 pixels is required for strong local coupling to occur. During the sensory stage each particle changes its orientation to rotate (via the parameter rotation angle, RA) towards the strongest local source of chemoattractant (Fig.~\ref{particlemorphology}b).  After the sensory stage, each particle executes the motor stage and attempts to move forwards in its current orientation (an angle from 0-360$^\circ$) by a single pixel forwards.  Each lattice site may only store a single particle and-critically-particles deposit chemoattractant into the lattice only in the event of a successful forwards movement (Fig.~\ref{particlemotorbehaviour}a). If the next chosen site is already occupied by another particle the default (non- oscillatory) behaviour is to abandon the move, remain in the current position, and select a new random direction (Fig.~\ref{particlemotorbehaviour}b). Diffusion of the collective chemoattractant signal is achieved via a simple 3x3 mean filter kernel with a damping parameter (set to 0.07) to limit the diffusion distance of the chemoattractant.

The low level particle interactions result in complex pattern formation. The population spontaneously forms dynamic transport networks showing complex evolution and quasi-physical emergent properties, including closure of network lacunae, apparent surface tension effects and network minimisation.  An exploration of the possible patterning parameterisation was presented in \cite{jones2010characteristics}. Although the particle model is able to reproduce many of the network based behaviours seen in the \Phys  plasmodium such as spontaneous network formation, shuttle streaming and network minimisation, the default behaviour does not exhibit oscillatory phenomena and inertial surging movement, as seen in the organism. This is because the default action when a particle is blocked (i.e. when the chosen site is already occupied) is to randomly select a new orientation-resulting in very fluid network evolution, resembling the relaxation evolution of soap films, and the lipid nanotube networks seen in \cite{lobovkina2008shape}.

The oscillatory phenomena seen in the plasmodium are thought to be linked to the spontaneous assembly / disassembly of actin- myosin and cytoskeletal filament structures within the plasmodium which generate contractile forces on the protoplasm within the plasmodium. The resulting shifts between gel and sol phases prevent (gel phase) and promote (sol phase) cytoplasmic streaming within the plasmodium. To mimic this behaviour in the particle model requires only a simple change to the motor stage. Instead of randomly selecting a new direction if a move forward is blocked, the particle increments separate internal co- ordinates until the nearest cell directly in front of the particle is free.  When a cell becomes free, the particle occupies this new cell and deposits chemoattractant into the lattice (Fig.~\ref{particlemotorbehaviour}c). The effect of this behaviour is to remove the fluidity of the default movement of the population. The result is a surging, inertial pattern of movement, dependent on population density (the population density specifies the initial amount of free movement within the population). The strength of the inertial effect can be damped by a parameter (pID) which sets the probability of a particle resetting its internal position coordinates, lower values providing stronger inertial movement.

When this simple change in motor behaviour is initiated surging movements are seen and oscillatory domains of chemoattractant flux spontaneously appear within the virtual plasmodium showing characteristic behaviours:  temporary blockages of particles (gel phase) collapse into sudden localised movement (solation) and vice versa. The oscillatory domains them- selves undergo complex evolution including competition, phase changes and entrainment. We utilise these dynamics below to investigate the possibility of generating useful patterns of regular oscillations which may be coupled to provide motive force.

\subsection{Model Setup}

The emergence of oscillatory behaviour in the model corresponds to differences in distribution of protoplasm within the plasmodium and subsequent changes in thickness of the plasmodium. In a real \Phys  plasmodium the changes in thickness of the plasmodium membrane are used to provide impetus (pumping of material through the vein network, or bulk movement of the plasmodium). There is known to be a relationship between the spontaneous contraction of the plasmodium and the subsequent transport of protoplasm away from that region. Thus the region undergoing contraction becomes thinner (allowing more light to pass through when illuminated) and regions away from the contraction become thicker as more protoplasm in present (allowing less light to pass through). In the computational model the transport of particles represents the free flux of protoplasm within the material and the increase in flux (mass particle movement) is indicated in the supplementary video recordings by an increase in greyscale brightness (http://uncomp.uwe.ac.uk/jeff/engines.htm). A decrease in the bulk movement of particles represents congestion and a lack of transport and is indicated by a decrease in greyscale brightness (since deposition of chemoattractant factor only occurs in the event of successful forward movement). For clarity in the static images, the greyscale images are inverted (dark areas indicate greater flux). 

The particle population's environment is a 2D lattice, represented by a digitised image configured to represent the habitat of the experimental plasmodium. We designed simple shapes in which the particle collective, which composes the virtual material, is confined. The particle population is free to move within unconfined areas. The shapes are composed of 'wall' regions where movement cannot occur, 'vacant' regions where movement was possible and (where relevant) 'stimulus' regions which provide attraction stimuli, or repulsion stimuli, to the particle population. At the start of each experiment the particle population is randomly distributed through the vacant space in the experimental arena and all particles have random initial orientations. A fixed population size was used (we do not discuss an adaptive population size in this report). The total amount of free possible particle movement is dependent on the population size as a fraction of vacant space. In these results we use a 90\% occupancy rate unless otherwise specified, i.e. 90\% of all vacant areas are occupied by particles. In all of the experiments there is an initial period where oscillatory behaviour is not initially 'switched on' and this results in self-organised regular domains. When oscillatory motor behaviour is induced these regular domains collapse and the emergence of small domains of regular oscillatory patterns begins. Over time these domains coalesce and compete, causing entrainment of the population into regular oscillation patterns, influenced by both the particle sensory parameters and also by the shape of the experimental arenas.

We use the arena shape to constrain the oscillatory behaviour to ascertain the possibility of utilising the oscillatory behaviours to provide useful 'engine-like' output. By this we mean that the oscillation patterns should be periodic, regular and reliable, in the same way that mechanical engines provide regularly and reliably timed patterns of impetus.

\subsection{Data Analysis}

We sampled regular frames from the emergent chemoattractant flux patterns and analysed the differences in particle flux by comparing the greyscale levels in different regions of the arena. We were particularly concerned with the reliable initiation of oscillatory behaviour and the characteristics of the behaviour (e.g. the period, the intensity and any coupling effects).

\section{Results}\label{sec:results}

\begin{figure}[tbp!]
 \begin{center}
  \includegraphics[width=0.85\textwidth]{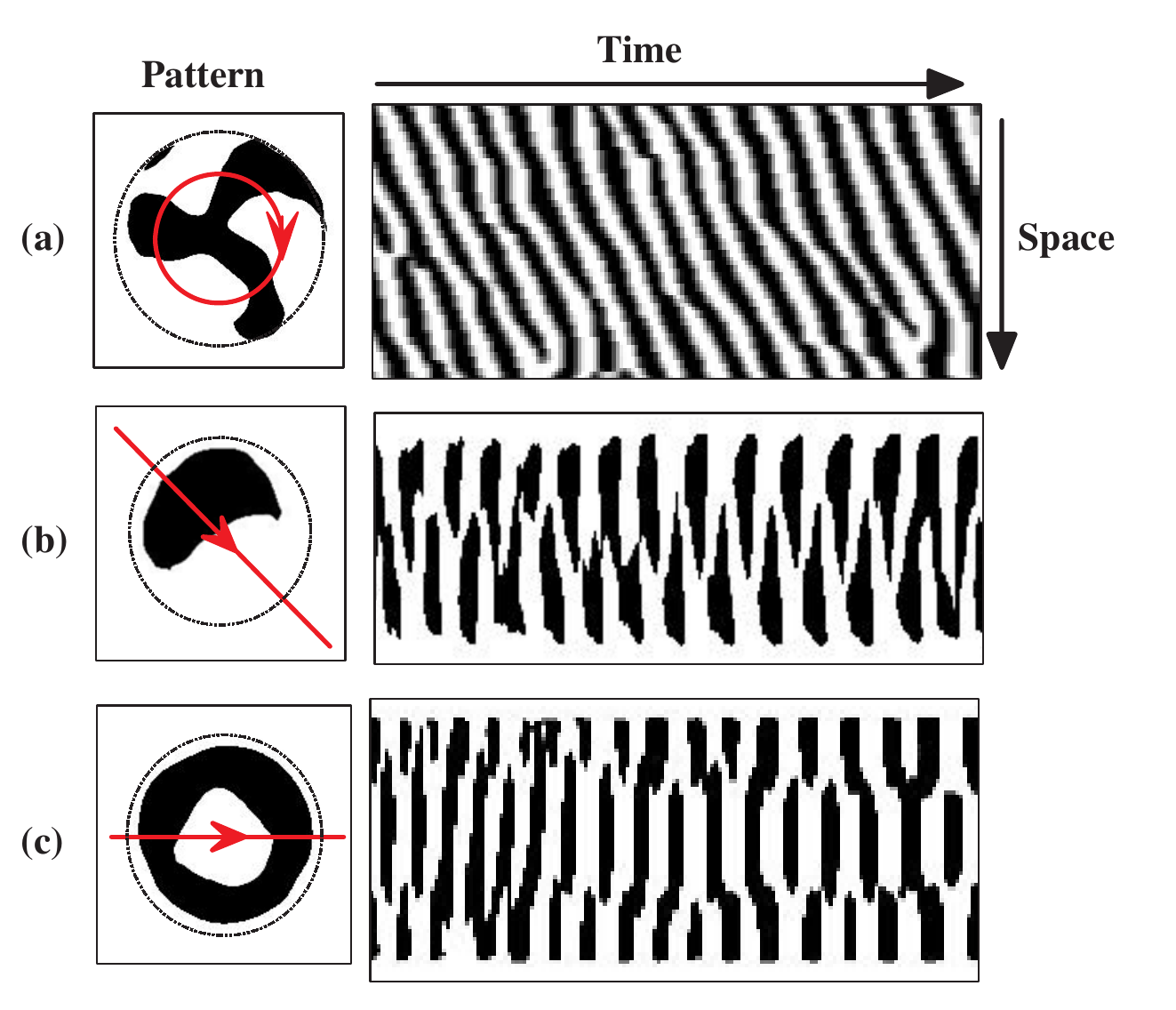}
 \end{center}
 \caption{Oscillation patterns observed in a particle model of {\itshape Physarum polycephalum} constrained within a circular well.
Left: Snapshot of pattern with trajectory of space time plot overlaid, Right: Space time plot of oscillation pattern.
a) Rotational pattern, b) Bilateral pattern, c) Annular pattern} 
 \label{particlewellpatterns}
\end{figure}

We have previously shown that the particle population successfully reproduces the behaviour of the plasmodium when confined within a circular well, resulting in the spontaneous emergence of oscillation patterns (summarised in (Fig.~\ref{particlewellpatterns}). The type of pattern produced depends on the sensory parameters of each particle (SA and RA parameters) and the SO parameter which, when increased, causes a transition to a different pattern (see \cite{TsudaS10PhysarumOsciAlife} for more information). In the circular well the patterns most frequently observed were rotational patterns. When the SO interaction distance was further increased lateral oscillations were observed, followed by an annular pattern. There appears to be a relationship between the circular confining shape of the arena and the type of pattern produced.

We speculate that the reason for the change in pattern type (which, over time, is also observed in the real plasmodium) is that the scale of interactions between the oscillatory regions becomes too large for the circular well, the interactions are constrained and a new pattern is formed which can 'fit' within the confines of the well. This suggests that the intrinsic oscillations which spontaneously emerge can be shaped in some way by architectural changes to the environment. To further explore this relationship between oscillation pattern type and the environment shape we explored different methods of patterning the virtual arena in order to assess differences in oscillation pattern type and pattern evolution.

\subsection{Transport Motion in Open Ended Patterns}

\begin{figure}[tbp!]
 \begin{center}
  \includegraphics[width=0.85\textwidth]{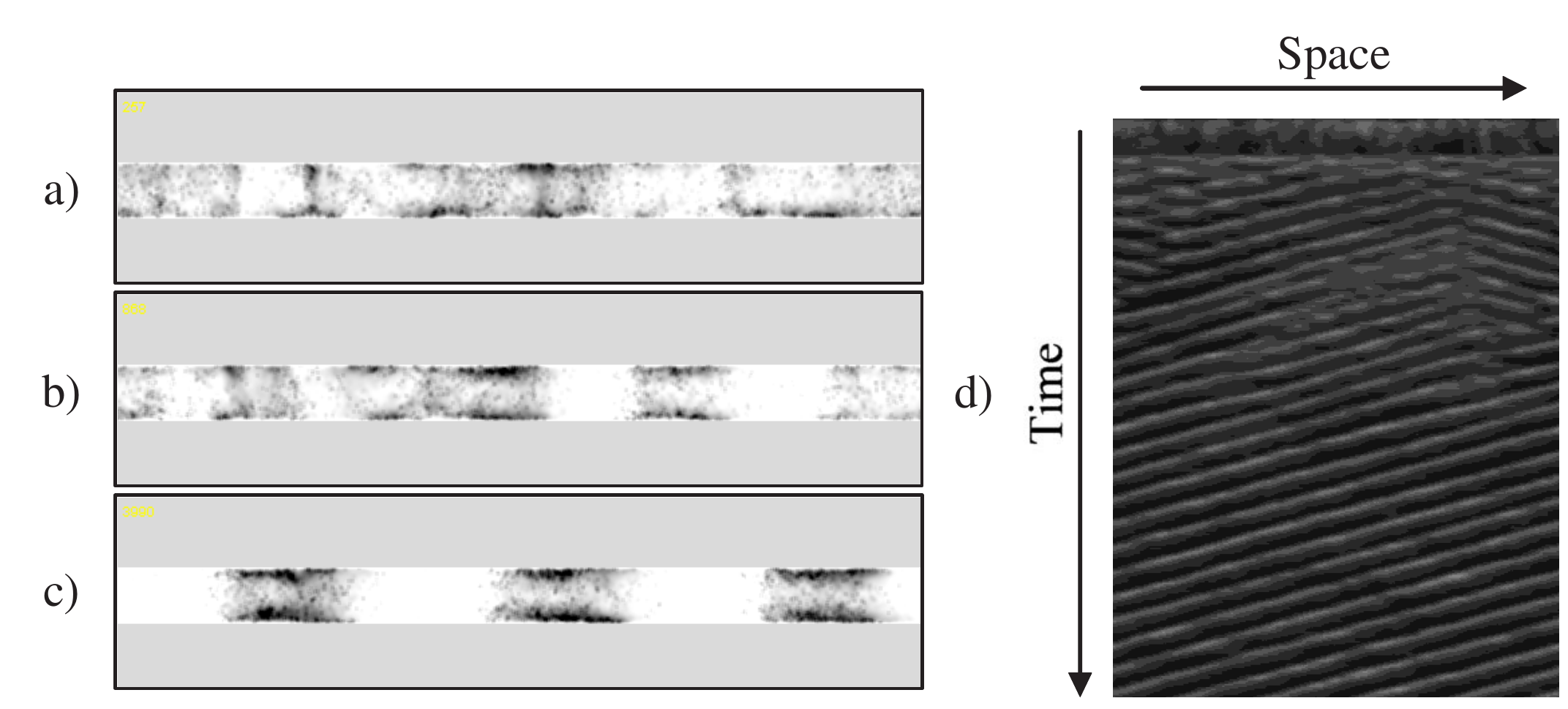}
 \end{center}
 \caption{Emergence of travelling wave transport in a tube shaped environment
Left: Snapshots taken at a) 257, b) 868 and c) 3990 scheduler steps
Right: Space time plot showing emergence of regular transport} 
 \label{linearopen}
\end{figure}

We patterned the environment with a simple tube shape whose ends were looped around at the edge of the environment by invoking periodic boundary conditions for the particles. The wall boundaries confined the population to the tube. Snapshots and a space-time plot are shown in Fig.~\ref{linearopen}. The space time plot is recorded by sampling every pixel along the width of the image at exactly halfway down the tube section and assembling an image based upon these sampled values (pixel brightness is related to chemoattractant trail concentration). After a short period where the non-oscillatory motor condition was used (see top of space-time plot) the oscillatory motor behaviour was initiated. Small domains representing different concentrations of chemoattractant flux appeared (Fig.~\ref{linearopen}a) which were travelling in different directions. Over time these domains competed and became fewer in number (Fig.~\ref{linearopen}b) until the tube was evenly divided into regular domains of high flux (regions of free movement) and low flux (regions of obstructed movement). These self-organised domains travelled in a single direction in a regular manner (Fig.~\ref{linearopen}c and lower section of space-time plot) and are an example of a phase transition into ordered movement patterns as first described in \cite{vicsek1995novel}.

\begin{figure}[tbp!]
 \begin{center}
  \includegraphics[width=0.85\textwidth]{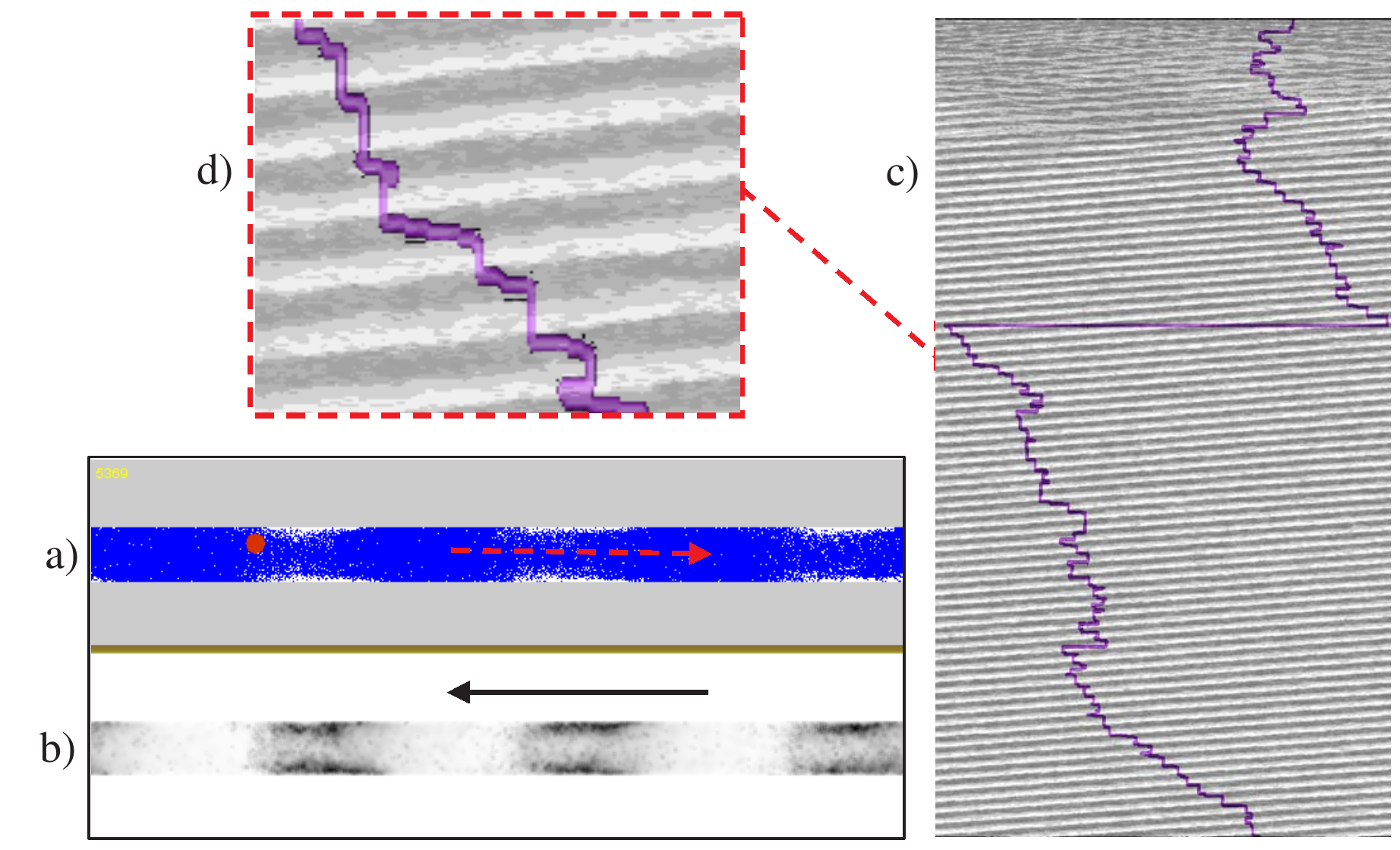}
 \end{center}
 \caption{Collective particle drift is opposite to the direction of wave propagation. a) Particle positions and tracer particle (circled), b) Chemoattractant wave propagation, c) Space-time plot overlaid with tracer particle position, d) Enlarged portion showing tracer particle movement (horizontal movement) tends to occur in vacant (high chemoattractant flux) areas.} 
\label{linearopendrift}
\end{figure}

The travelling waves arising from the collective particle behaviour actually travel in the opposite direction to the bulk particle motion. In the example shown in Fig.~\ref{linearopendrift} the direction of the travelling wave (Fig.~\ref{linearopendrift}b, solid arrow) is right-to-left, whereas the actual movement of the particles (Fig.~\ref{linearopendrift}a, dashed arrow) tends to move from left-to-right. Since particles can only move when a space in the lattice becomes available the vacant spaces appear to move in the opposite direction of the particles and the particles themselves show a greater probability of moving when a region of relatively vacant space is near (Fig.~\ref{linearopendrift}c, also note that particle movement tends to occur in the lighter regions, which are regions with more vacant space and greater chemoattractant flux). The spaces themselves appear to move backwards because a particle moving from its current position to occupy a vacant space subsequently leaves a new vacant space at its previous position. Because chemoattractant is only deposited by the particles after a successful movement, the chemoattractant flux will be patterned by the distribution of vacant spaces and the wavefront thus moves in the opposite direction to the particles. The bulk movement of the particles is also much slower than the travelling waves, as indicated by Fig.~\ref{linearopendrift}, which shows that a single particle takes approximately 8000 scheduler steps to traverse the width of the arena, whereas the travelling wave crosses the arena in approximately 400 steps, a 20:1 difference (although the particle's progress is hindered somewhat by the resistance caused by the low numbers of vacant spaces and changes in direction on contact with the border regions). The opposite direction of the self-organised travelling wave with respect to particle movement is reminiscent of the characteristic backwards propagation seen, for example, in traffic jams \cite{flynn2009jamitons}.

\begin{figure}[tbp!]
 \begin{center}
  \includegraphics[width=0.85\textwidth]{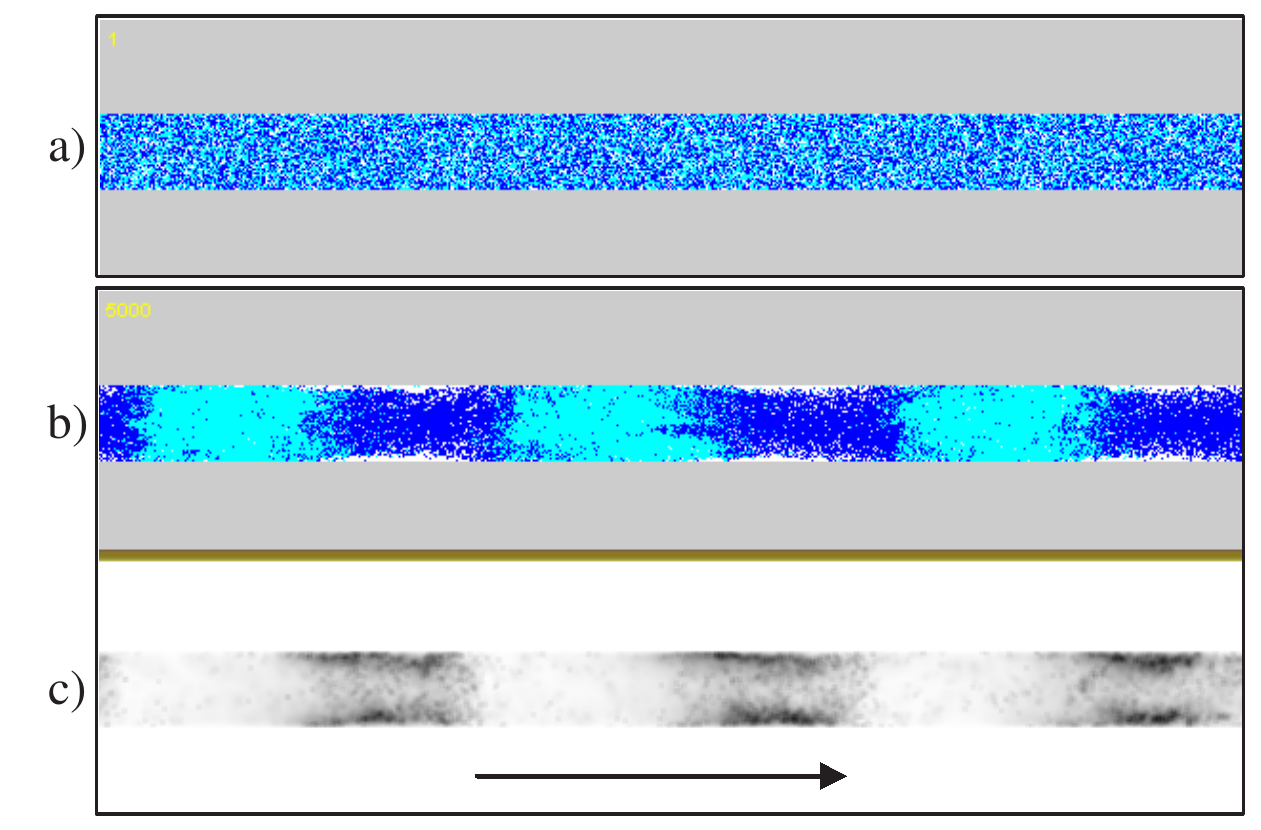}
 \end{center}
 \caption{Collective particle alignment and the travelling wave. a) Initial random distribution of particles and their alignments, b) When the stable travelling wave occurs particles are distributed into groups sharing similar directional alignment: lighter regions are oriented to face right and darker regions are oriented leftwards, c) Chemoattractant flux in travelling wave moving to the right.} 
\label{linearopenangledist}
\end{figure}

Further analysis of the directional alignment of the particles revealed some unexpected properties of the particle population in relation to the travelling wave. At the start of an experiment the distribution and alignment of the particle population is randomly chosen (Fig.~\ref{linearopenangledist}a) and the ratio of particles facing left and those facing right is typically equal at 50:50 (we consider particle alignment as being 'left facing' or 'right facing', depending on whether their actual orientation angle falls into either category). When the travelling wave has emerged and stabilised the distribution of the particles is such that they are grouped in regions (Fig.~\ref{linearopenangledist}b) which are similar - but slightly offset - to the regions of high and low chemoattractant concentration seen in the travelling wave (Fig.~\ref{linearopenangledist}c). The particles in each region share the same general alignment and the different regions alternate with respect to the alignment of the particles within them (i.e. regions are LEFT, RIGHT, LEFT, RIGHT and so on). The final alignment ratio at the end of an experiment was typically 53:47 (rounded average of ratios over ten runs), with the majority of particles facing in the direction of the travelling wave. The particles change their directional alignment at the same speed as the oncoming travelling wave. The actual movement of the particles, however, is much slower and in the opposite direction to the wave.

\begin{figure}[tbp!]
 \centering
 \subfigure[]{\includegraphics[width=0.6\textwidth]{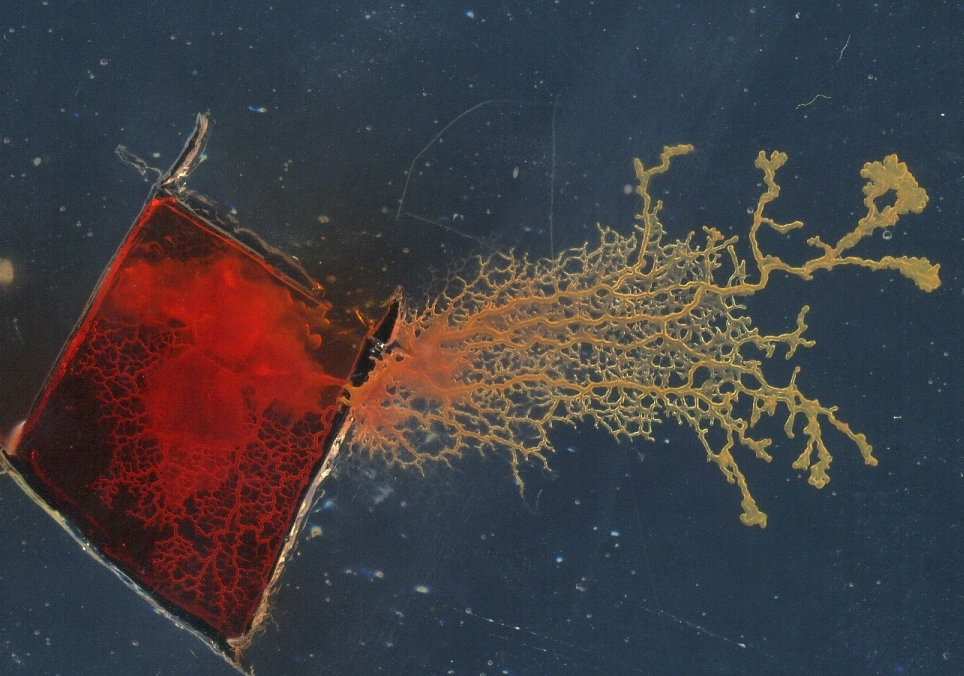}}
 \subfigure[]{\includegraphics[width=0.6\textwidth]{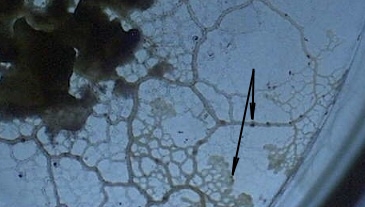}}
 \caption{Evidences of \Phys transport: (a)~Propagating plasmodial localisation transports red dye outside the inoculation site. See details in~\cite{AdamatzkyPainting}.  (b)~Protoplasmic network with clusters of silver spheres transported away from their source. Examples of  two clusters are marked by arrows.} 
\label{physsilver}
\end{figure}

Contractile waves travelling along protoplasmic tubes (Fig.~\ref{linearopenangledist}) can be used to pump liquid substances and, quite likely, to transport fine-grained granular substance. Two examples are shown in Fig.~\ref{physsilver}, see detailed explanation of experimental setups in~\cite{AdamatzkyPhysarumMachines}.

First example demonstrates pumping of red food colouring along tubes of the growing protoplasmic tree. Initially, a piece of plasmodium was inoculated in the small domain of agar physically insulated from the rest of agar plate. An oat flake saturated with red food colouring is placed nearby the plasmodium, also inside the insulated domain. The plasmodium propagates outside the domain and crosses the agar-less boundary. It pumps the food colouring across the boundary Fig.~\ref{physsilver}a~\cite{AdamatzkyPainting,AdamatzkyPhysarumMachines}.

Second example deals with transport of granular substances. We mixed an oat flake powder with  
a small quantity of hollow silver-coated spheres (10-27 $\mu$m diameter, Cospheric Inc, Santa Barbara, CA) and 
inoculated \Phys on top of the mixture. Preliminary experiments suggest that the spheres are transported both inside the protoplasmic tubes (usually solitary spheres) and on the outside of the tube's membrane (usually larger conglomerates of spheres)
(Fig.~\ref{physsilver}a).

\begin{figure}[tbp!]
 \begin{center}
  \includegraphics[width=0.85\textwidth]{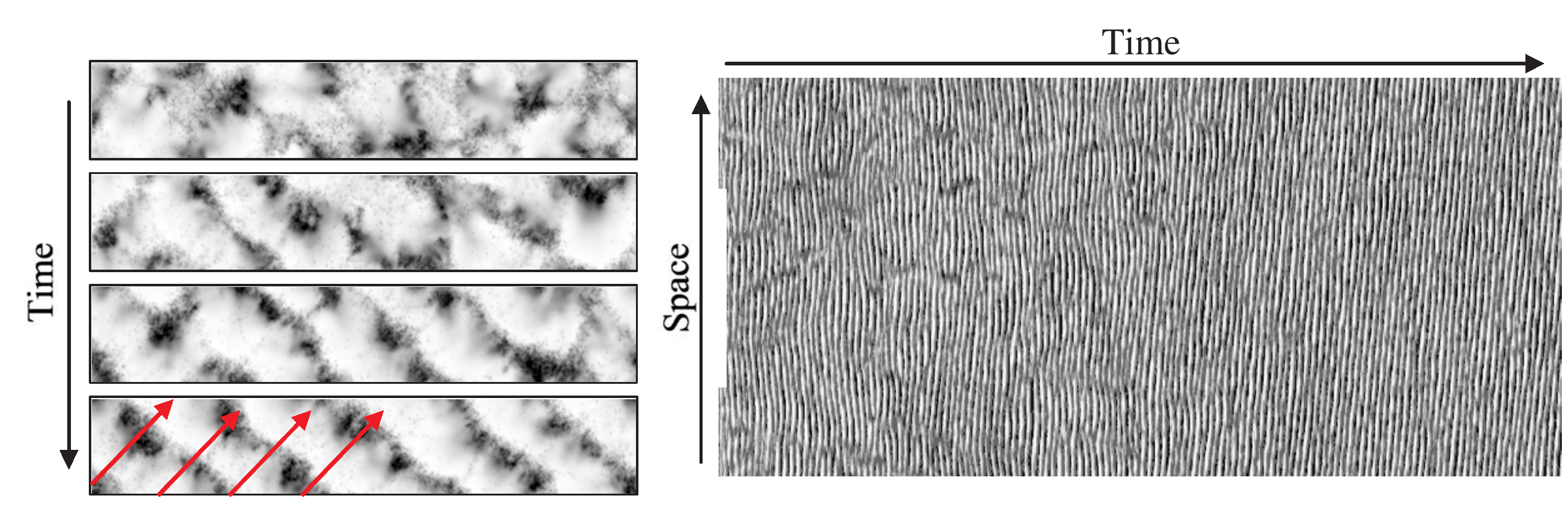}
 \end{center}
 \caption{Helical transport in non-patterned environment with periodic boundary conditions. Left: Snapshots of competition and entrainment of wavefronts with final helical-type movement arrowed, Right: Space-time plot showing long period before regular transport occurs.} 
\label{helical}
\end{figure}

By patterning the environment to remove all wall boundaries and ensure periodic boundary conditions the emergent oscillatory patterns self-organised into travelling waves (Fig.~\ref{helical}). However there was a much greater length of time needed for the competition between the wave patterns to complete and form synchronous travelling waves. The increase in time before synchronous waves emerge can be explained by the greater initial freedom of movement afforded by the lack of movement constraints from the environment.

\begin{figure}[tbp!]
 \begin{center}
  \includegraphics[width=0.85\textwidth]{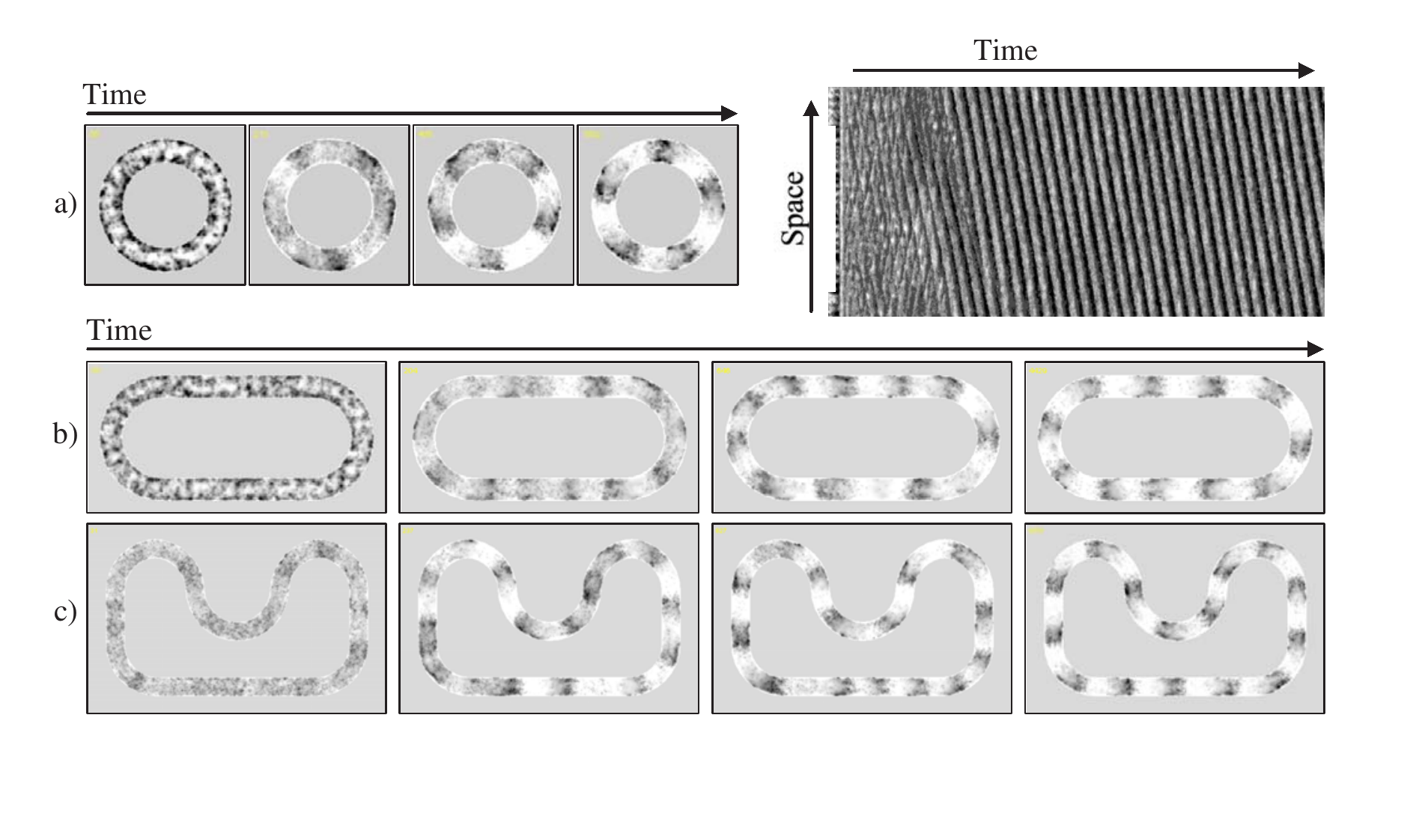}
 \end{center}
 \caption{Emergence of rotary motion oscillations. a) Emergence of rotary motion and visualisation of space-time plot showing regular motor pulses. Space time plot was created by recording 360 points inside the vacant track of the circle, b) Tracked rotary movement from a combination of circular and straight regions, c) Conveyor type motion from a more tortuous looped structure.} 
\label{rotaryrings}
\end{figure}

By removing the movement at the boundaries and instead patterning the vacant space into looped structures rotary motion of the travelling waves was achieved (Fig.~\ref{rotaryrings}). The competition period before synchronisation was relatively brief (Fig.~\ref{rotaryrings}a, space-time plot), again because the environmental barriers reduce initial freedom of movement. The looped structures also enabled travelling wave motion even when the environment added non-circular elements, and more tortuous paths (Fig.~\ref{rotaryrings} b and c). The successful initiation of rotary motion suggests the possibility of generating reliable conveyor type transport.

\begin{figure}[tbp!]
 \begin{center}
  \includegraphics[width=0.85\textwidth]{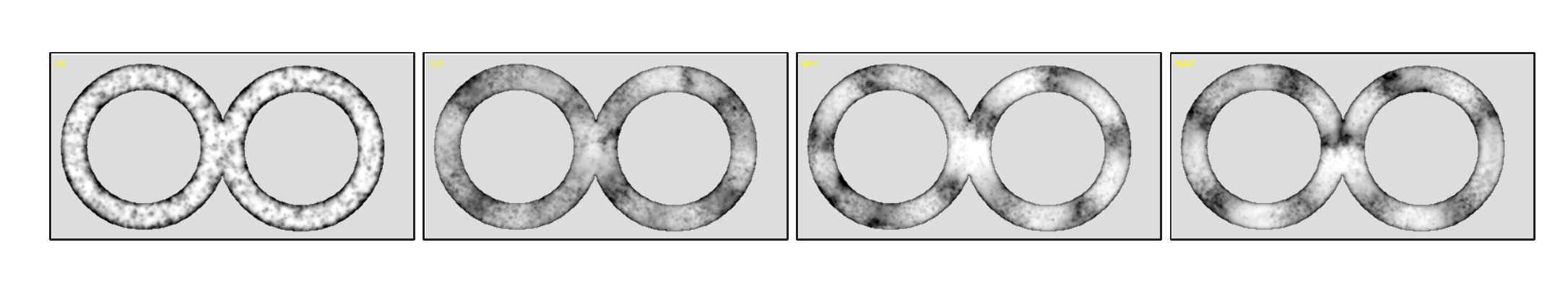}
 \end{center}
 \caption{Emergence of gear-like coupling in an overlapping two rotor pattern. Snapshots taken at 20, 121, 981 and 6997 scheduler steps.} 
\label{gears}
\end{figure}

By combining two circular patterns and introducing a region which overlapped and exposed to two separate wavefronts it was possible to achieve entrainment, after 5000 scheduler steps, of the pulses from one rotary 'motor' to another, mimicking the transmission and synchronisation of movement to another ring by a fluidic coupling effect (Fig.~\ref{gears}). Like a conventional gear transmission, the rotation of two facing rings was in opposite directions, however, unlike conventional gear trains, the 'teeth' (wavefront peaks) did not overlap.

\begin{figure}[tbp!]
 \centering \subfigure[]{\includegraphics[width=0.5\textwidth]{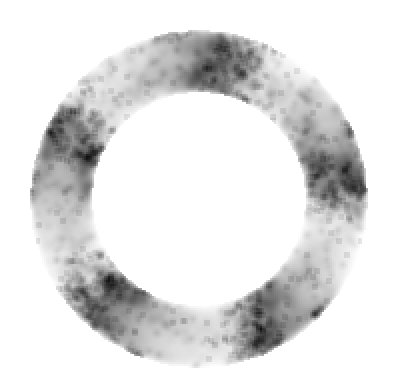}} \subfigure[]{\includegraphics[width=0.5\textwidth]{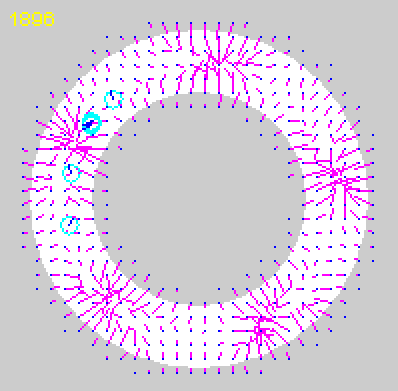}}
 \caption{Simulation of passive transport of substances using information from travelling waves. (a)~Spatial representation of snapshot of emergent travelling waves.  (b)~Computed vector field based upon direction and gradient of travelling waves is used to move passive objects (circular shapes).} 
\label{model_transport}
\end{figure}

To simulate the transport of substances using the emergent travelling waves of the model we computed a vector field based at every scheduler step on instantaneous snapshots of the wave activity (Fig.~\ref{model_transport}a). The vector field was computed by dividing the habitable areas of the lattice into 15$\times$15 pixel regions. Each vector value was located at the centre of a region and was computed from the difference of chemoattractant flux at extreme borders of each region. The difference between left and right borders giving the \textbf{dx} value and the difference between top and bottom borders giving the \textbf{dy} value. These values were converted to polar coordinates to provide the angle of the strongest local concentration (direction) and the force given to the transported substances. The values within the computed vector field were used to move passive substances, whose locations are represented by the circular areas (Fig.~\ref{model_transport}b). The supplementary video recording illustrates how information within the travelling wave patterns can be translated into motive force to transport the passive substances. The recording of the vector field also shows the direction and size of the inputs to the transported substances.

\subsection{Transport in Closed Path Patterns}

\begin{figure}[tbp!]
 \begin{center}
  \includegraphics[width=0.85\textwidth]{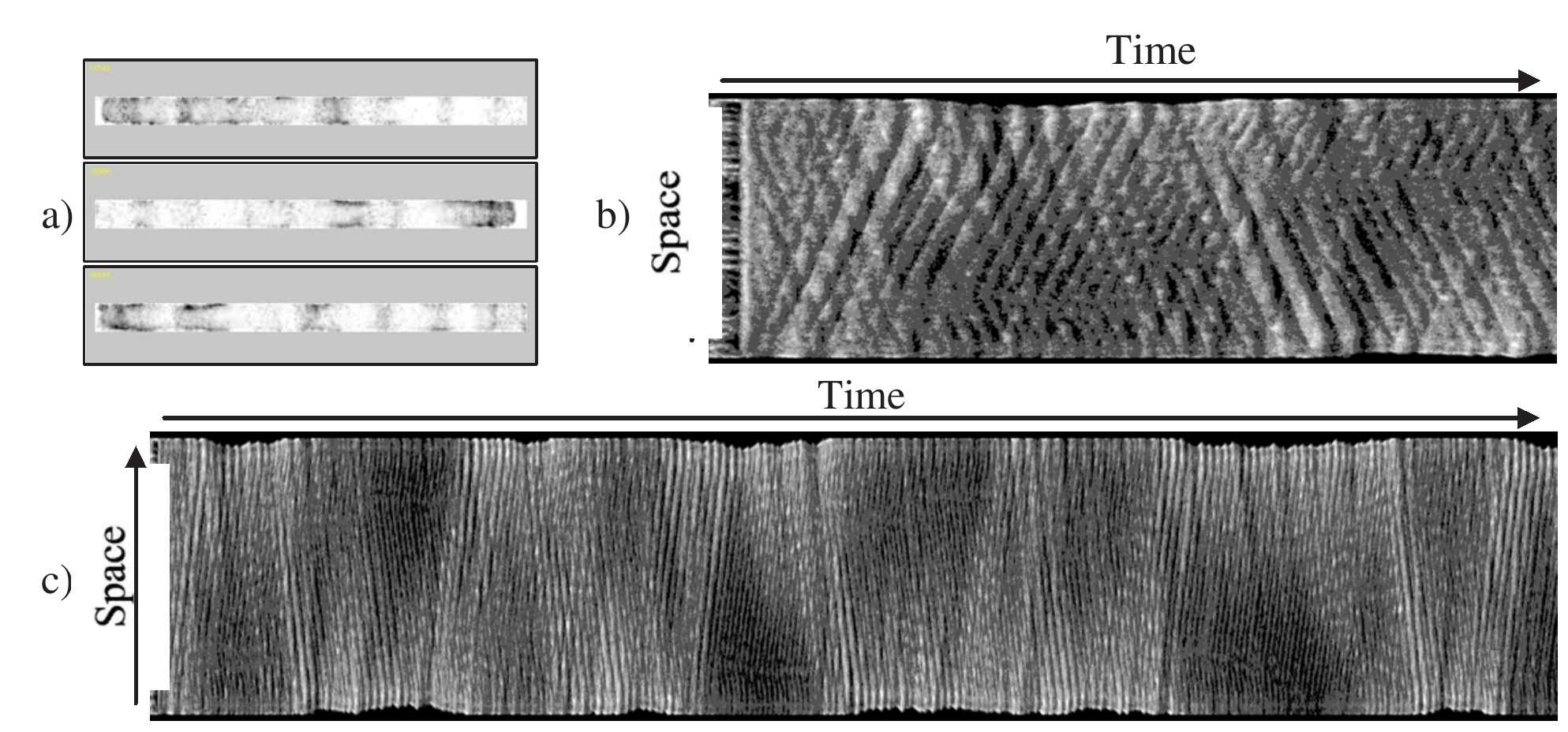}
 \end{center}
 \caption{Complex second-order oscillations caused by bulk drift in closed path environments. a) Pulsatile oscillations at different sides of a closed chamber, b) Initial phase of space-time plot showing initiation of oscillatory behaviour and competition between oscillatory domains, c) long term (20,000 steps) space-time plot showing second order oscillations as bulk particle positions oscillate from one side of the chamber to the other. Dark regions at either side of plot indicate periods of second-order oscillations.} 
\label{closedtube}
\end{figure}

The transport motion in open ended looped patterns stabilises because the bulk particle drift eventually synchronises with the travelling waves. The distribution of the particle population becomes relatively evenly distributed within the path, punctuated by regularly spaced changes of particle occupation density. When closed path patterns were used, however, the uniform distribution cannot occur because separate ends of the path cannot communicate the transport of particles. Thus, over time, the drift of particles results in a tendency for the particle population density to become greater at one end of the chamber. Once there is an imbalance at one end the number of vacant spaces at that end falls and the particles are then attracted to the opposite side of the chamber where more vacant areas exist. Short term oscillatory transport and competition within the chamber still occurs (Fig.~\ref{closedtube}a and b), as with looped patterns, but this is mediated by a second order of oscillations, which occurs over a much longer timescale since it is caused by the slower bulk drift of particles (Fig.~\ref{closedtube}c). The effect is regular changes in direction of transport direction and this suggests a possible mechanism of how changes in direction in open-ended looped systems could be achieved - by temporarily introducing a blockage in a looped path until a change in direction occurred.

\begin{figure}[tbp!]
 \begin{center}
  \includegraphics[width=0.99\textwidth]{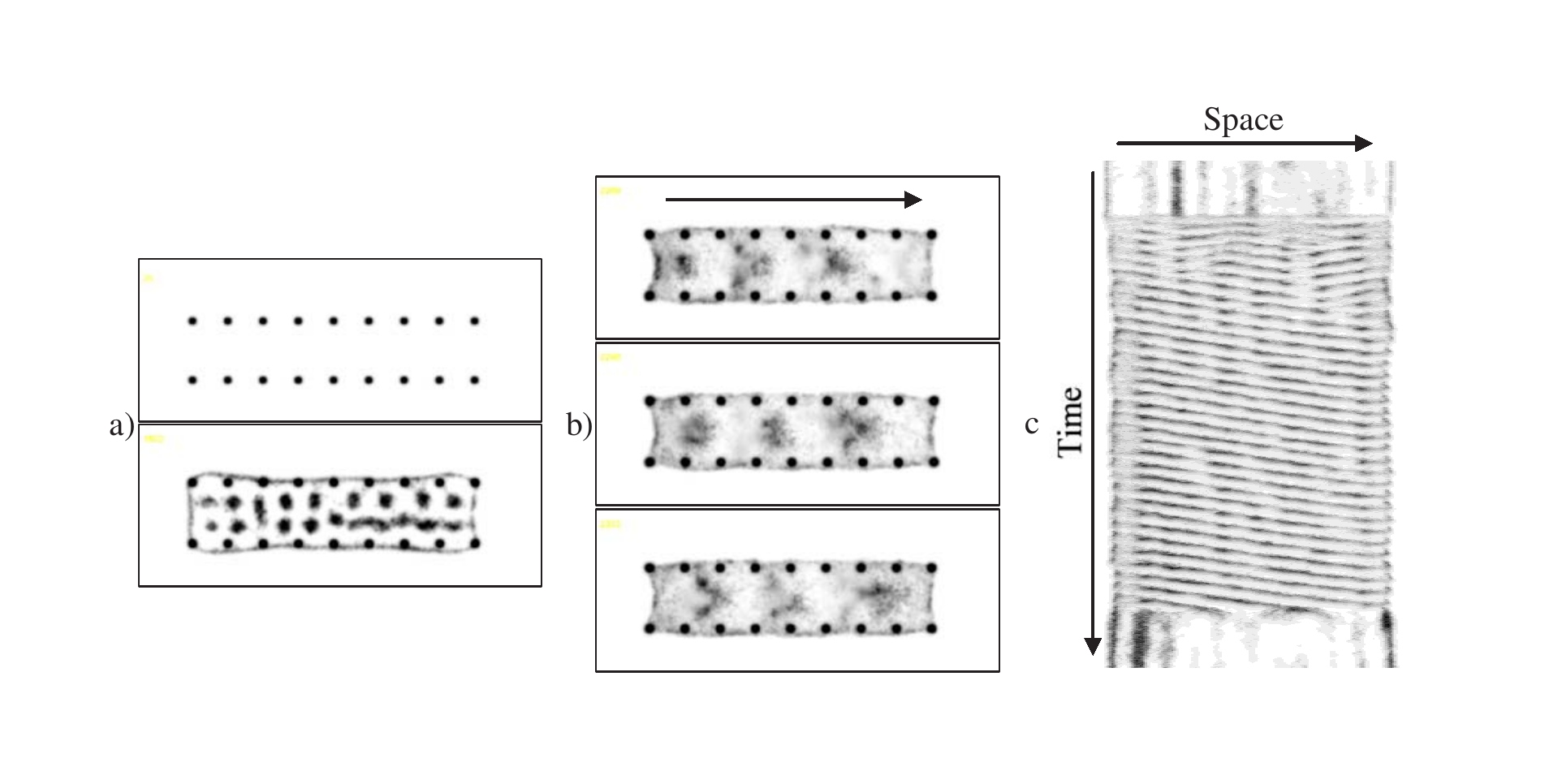}
 \end{center}
 \caption{Confining the collective by attractant projection and emergence of travelling wave. a) Regularly spaced projection of chemoattractant sources (top) confines position of collective (bottom), b) When oscillatory motor behaviour is activated a travelling wave emerges across the confined collective, c) Space-time plot of the emergence of travelling waves in a pinned collective.
} 
\label{pinned}
\end{figure}

Another method to confine the collective to a region is by using strong sources of attractant to effectively 'pin down' the collective in place instead of confining it physically within a region by its boundaries.
Attractant sources are represented by projecting values at every scheduler step into the chemoattractant diffusion field. These sources diffuse and attract the individual particles of the collective. By spacing the placement of attractants the collective is retained in place in a sheet-like fashion by the attraction of the particles to the sources and the mutual attraction of particles to their own deposition of chemoattractant into the diffusion field (Fig.~\ref{pinned}a).

When oscillatory motor behaviour is activated travelling waves of chemoattractant movement emerge in different directions (Fig.~\ref{pinned}b). These waves compete for a short time before one direction predominates (Fig.~\ref{pinned}c).

\section{Conclusion and Discussion}\label{sec:discussion}

We have examined the problem of generating, controlling and distributing motive forces at very small scales using the true slime mould {\itshape Physarum polycephalum} as a prototype mechanism. \Phys is attractive because it
satisfies many physical and computational properties which are desirable
in robotics applications (self-oscillatory, simple components,
distributed sensory and motor control, integration of multiple sensory
stimuli, amorphous and adaptive shape, amenable to external influence,
resilience to damage, self repair). Using the organism itself we have
demonstrated and measured its ability to lift and pump material using
its intrinsic oscillations and protoplasmic streaming. We also
investigated mechanisms by which the pumping and transport behaviour may
be externally influenced and inhibited, using irradiation stimuli, so
that it may be used as a prototype steering mechanism. We used actual
experimental data of the bilateral output from a dumbbell shaped
plasmodium to drive a model mechanism, a Braitenberg-type vehicle and
used the irradiation stimulus to act as a steering force. 

In the experiments presented here, the size of {\itshape Physarum} cells
was about a few millimeter scale. However, the possible minimum size of
the {\itshape physarum} plasmodium is known to be approximately tens of
microns. Considering the size-invariant oscillating behaviour of the
{\itshape Physarum} slime mould, it should be possible to construct a
micron-scale {\itshape Physarum} engine to actuate tiny objects.

Moving beyond the use of the plasmodium as a single engine, or pumping, element, we explored how the distributed transport within the organism could be reproduced for synthetic distributed transport. By patterning a densely packed particle population into different shapes we were able to simulate reliable and regular oscillatory movement as emergent travelling waves of plasmodial contractions which were fashioned into rotary, reciprocal, helical, and coupled transport mechanisms. The waves, consisting of peaks and troughs of simulated chemoattractant flux, were found to travel at opposite directions, and with much greater velocity, than the underlying particle drift motion which generated the waves. By computing a simple vector field from the wave motion it was possible to simulate motive forces to transport substances within the environment. 

{\itshape Physarum} slime mould is an organism which efficiently exploits self-oscillatory phenomena to transport nutrients within its protoplasmic network and can inspire material approaches to distributed movement and transport. We envisage that these prototype results of our studies on slime-mould engines and transporting devices could be applied in the design and manufacturing of intelligent soft manipulator arrays in micro-electronic assembly, drug delivery systems and novel prototypes of artificial tissue and organs.

\section{Acknowledgement} 

The work was partially supported by the Leverhulme Trust research grant
F/00577/1 ``Mould intelligence: biological amorphous
robots''. Experiments in Section~\ref{sec:force-gener-itsh} was
conducted in Prof Hywel Morgan's lab at University of Southampton as a
part of the project supported by the Life Sciences Interfaces Forum
(Project manager: Dr Klaus-Peter Zauner).


\end{document}